\documentclass{aa}
\usepackage{graphicx}
\usepackage{amssymb }
\begin{document}
   \title{The very low-mass population of the Corona Australis and
   Chamaeleon~II star forming regions
     \thanks{Based on observations collected at the European Southern 
       Observatory, La Silla, Chile}
   }

   \subtitle{}

   \author{Bel\'en L\'opez Mart\'{\i},\inst{1}
 	  Jochen Eisl\"offel\inst{2}
	  \and Reinhard Mundt\inst{3}
          }

   \offprints{B. L\'opez Mart\'{\i}}

   \institute{Departament d'Astronomia i Meteorologia, Universitat de Barcelona,
   	Mart\'{\i} i Franqu\`es 1, E-08028 Barcelona, Spain\\
              \email{blopez@am.ub.es}
         \and
	 Th\"uringer Landessternwarte, Sternwarte 5, D-07778 Tautenburg,
   	 Germany\\
              \email{jochen@tls-tautenburg.de}
		\and Max-Planck-Institut f\"ur Astronomie, K\"onigstuhl 17, 
		D-69117 Heidelberg, Germany\\
		\email{mundt@mpia-hd.mpg.de}
            }

   \date{Received ; accepted }

   \abstract{We present the results of a deep optical survey in the Corona
   Australis and Chamaeleon~II star forming regions. Our optical photometry is
   combined with available near- and mid-infrared photometry to identify very
   low-mass candidate members in these dark clouds. In our Chamaeleon~II field,
   only one object exhibits clear H$\alpha$ emission, but the discrepancy
   between its optical and near-infrared colours suggests that it might be a
   foreground star. We also identify two objects without H$\alpha$ emission
   that could be planetary mass members of Chamaeleon~II. In Corona Australis,
   we find ten stars and  three brown dwarf candidates in the Coronet cluster.
   Five of our new members are identified with ISOCAM sources.     Only two of
   them have a mid-infrared excess,  indicating the presence of an accretion
   disk. On the other hand, one brown dwarf candidate has a faint close
   companion, seen only in our deepest $I$-band image. For many of the
   candidates in both clouds, membership could not be inferred from their
   H$\alpha$ emission or near-infrared colours; these objects need
   spectroscopic confirmation of their status.

   }

\titlerunning{The VLM population of CrA and Cha~II}
\authorrunning{L\'opez Mart\'{\i} et al.}

   \maketitle

%########################################################################
\section{Introduction}\label{sec:introdcra}

	Surveys for very low-mass stars and brown dwarfs in star forming
regions and young clusters are of great importance to understand the formation
and early evolution of these objects, and to address the problem of the
universality of the initial mass function (IMF) in the low-mass regime. Many
such surveys have been carried out in the past years (e.g. B\'ejar et al.
\cite{bejar99}; Wilking et al. \cite{wilking99}; Barrado y Navascu\'es et al.
\cite{barrado01}; Mart\'{\i}n et al. \cite{martin01}; L\'opez Mart\'{\i} et al.
2004, 2005). In parallel, spectroscopic studies of young very low-mass objects
have been performed searching for signatures of accretion and mass loss (e.g.
Fern\'andez \& Comer\'on \cite{fernandez01}; Comer\'on et al. \cite{comeron03};
Jayawardhana et al. \cite{jay02}, \cite{jay03}; Scholz \& Eisl\"offel
\cite{scholz04};  Mohanty et al. \cite{mohanty05}),  and mid-infrared
observations have started to unveil the structure of circum(sub)stellar disks
(e.g. Natta \&  Testi \cite{natta01}; Testi et al. \cite{testi02}; Sterzik et
al. \cite{sterzik04};  Mohanty et al. \cite{mohanty04}).

	Since the observations show that young brown dwarfs and young very
low-mass stars have similar properties, it seems most likely that they form by
a similar process, namely the collapse and fragmentation of a molecular cloud
core. It is not yet clear, however, what mechanism prevents a brown dwarf core
from becoming a star. While some models try to explain the observed numbers of
substellar objects by turbulent collapse (e.g. Padoan \& Nordlund
\cite{padoan02}), other authors invoke the concurrence of an external mechanism
to stop the core growing process, such as winds and/or the ionizing radiation
from a nearby hot massive star (Whitworth \& Zinnecker \cite{whitworth04}), or
ejection from a multiple system (Reipurth \& Clarke \cite{reipurth01}; Bate et
al. \cite{bate03}; Delgado Donate et al. \cite{delgado03}).

	In this paper, we present the results from our optical multi-band
survey for very low-mass objects in two nearby star forming regions, the Corona
Australis and the Chamaeleon~II dark clouds. They are very suitable to
investigate the problem of low-mass star and brown dwarf formation because of
their proximity and high Galactic latitude, implying low contamination from
background objects. We followed the same method as in our previous surveys in
\object{Chamaeleon~I} (L\'opez Mart\'{\i} et al. 2004, hereafter \cite{lm04})
and \object{Lupus~3} (L\'opez Mart\'{\i} et al. 2005, hereafter \cite{lm05}).
The paper is structured as follows: First we summarize the characteristics of
our chosen regions in Sect.~\ref{sec:regions}. In Sect.~\ref{sec:wfi}, we
describe our observations, as well as the data reduction and photometry
procedures. The survey results are presented in Sect.~\ref{sec:bd}. The
properties of the newly identified cloud members are discussed in
Sect.~\ref{sec:cradisc} and \ref{sec:chaiidisc}. We  finish with the
conclusions in Sect.~\ref{sec:concl}.

%
%########################################################################
\section{Surveyed regions}\label{sec:regions}

%                                     
%______________________________________________ 
   \begin{figure}[t]
   \centering
   \includegraphics[width=8cm, angle=-90, bb= 20 50 550 500]{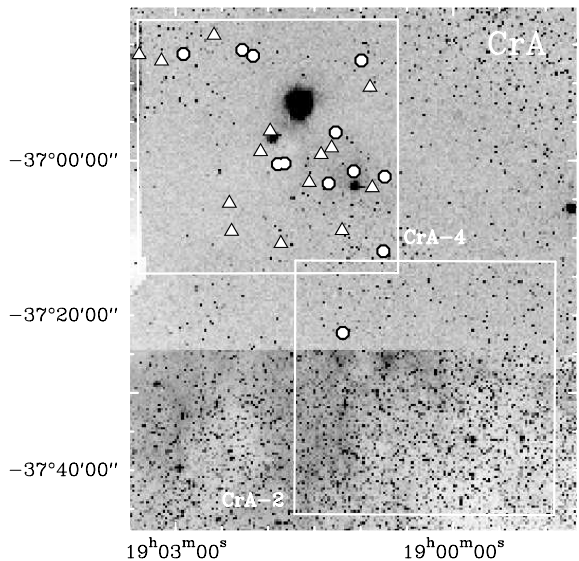}
    \caption{\footnotesize
		DSS image showing the location of our WFI fields (CrA-2 and -4)
		in Corona Australis. The circles and triangles indicate the
		positions of the objects with and without H$\alpha$ emission,
		respectively. The former ones are regarded as probable cloud
		members.
		}
              \label{fig:crafield}
    \end{figure}

\subsection{Corona Australis}\label{sec:cra}

	At a distance of 130~pc (Marraco \& Rydgren \cite{marraco81}), the 
\object{Corona Australis} molecular complex is a close and interesting site of
star  formation. Dominated by a centrally condensed core near the emission-line
star  \object{R~CrA}, this region is characterised by strong concentrations of
molecular  material (Loren \cite{loren79}). The cloud contains several bright
infrared  sources and Herbig-Haro objects (see Wang et al. \cite{wang04};
Graham \cite{graham91}, and references therein). Especially relevant was the
recognition of a number of heavily obscured low-mass stars in the R~CrA cloud
core making up a compact embedded cluster named ``the Coronet'' (Taylor \&
Storey  \cite{taylor84}).

	The young stellar population has been studied by means of H$\alpha$
surveys (Marraco \& Rydgren \cite{marraco81}), optical spectroscopy (Walter et 
al. \cite{walter97}), infrared mapping (e.g. Wilking et al. \cite{wilking92}) 
and X-ray observations (e.g. Patten \cite{patten98}; Neuh\"auser et al. 
\cite{neuhauser00}). Although no bona-fide brown dwarfs are yet known in Corona
Australis, deep infrared surveys have started to provide some good candidates 
(Wilking et al. \cite{wilking97}; Olofsson et al. \cite{olofsson99}). More
recently, Fern\'andez \& Comer\'on (\cite{fernandez01}) reported on an object 
in the transition from stars to brown dwarfs seen in the direction of the 
R~CrA core. This object, \object{LS-RCrA~1}, shows indications of intense
accretion and mass loss in its spectrum, and provided the first observational
evidence that such processes take place in young objects with masses down to
the substellar boundary (see also Barrado y Navascu\'es et al.
\cite{barrado04a}). Objects similar to LS-RCrA~1 have been later discovered in
other star forming regions (e.g. Comer\'on et al. \cite{comeron03}; Barrado y 
Navascu\'es \& Jayawardhana \cite{barrado04b}).

%______________________________________________________________
\subsection{Chamaeleon~II}\label{sec:chaii}

       The \object{Chamaeleon} star forming complex  consists of three dark
clouds, named Chamaeleon~I, II, and III. The largest and best studied of the
three is Chamaeleon~I: With an estimated age of about 3~Myr, it contains more
than 150 known young stars (e.g. Schwartz \cite{schwartz91}; Gauvin \& Strom
\cite{gauvin92}). This cloud has also been surveyed for brown dwarfs (e.g.
Comer\'on et al. \cite{comeron00}, \cite{comeron04}; Luhman \cite{luhman04}).
In particular, we identified 44 low-mass stars and 27 new brown dwarf
candidates in a WFI survey of an area of about 1.2 deg$^2$ in Chamaeleon~I
(\cite{lm04}). 

	\object{Chamaeleon~II} seems to be at an earlier stage of evolution
than Chamaeleon~I, because it contains more embedded than visual objects (e.g.
Gauvin \& Strom \cite{gauvin92}). This dark cloud has an extension of about
1.5~deg$^2$ and lies at an estimated distance of 178~pc (Whittet et al.
\cite{whittet97}). Surveys by several authors have identified at least 19
T~Tauri stars in this star forming region (Schwartz \cite{schwartz77}; Gauvin
\& Strom \cite{gauvin92}; Hughes \& Hartigan \cite{hughes92}). It also contains
an embedded Herbig Ae star, \object{IRAS~12496-7650} (Hughes et al.
\cite{hughes91}). More recently, Young et al. (\cite{young05}) identified more
than 40 potential young stellar objects in this cloud in a Spitzer-MIPS survey,
including two previously unknown sources with 24$\mu$m excesses.

	Up to now, no bona-fide brown dwarfs are known in Chamaeleon~II.
Recently, Vuong et al. (\cite{vuong01}) presented a list of 51 new objects in
this region from the DENIS survey with $I-J$ colours suggesting that they may
be very low-mass stars or brown dwarfs. Low- and mid-resolution spectra of some
of these objects were recently provided by Barrado y Navascu\'es \&
Jayawardhana (\cite{barrado04b}), who could only confirm one candidate,
\object{C41}, as a Chamaeleon~II member, with a mass near the substellar
boundary. This object also shows signatures of accretion and outflow in its
spectrum.

%______________________________________________ 
   \begin{figure}
   \centering
   \includegraphics[width=8cm, angle=-90, bb= 0 100 600 600]{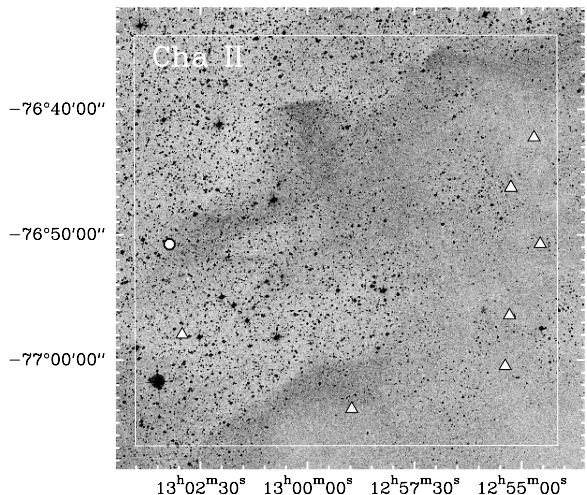}
    \caption{\footnotesize
		DSS image showing the positions of our objects in the surveyed
		WFI field in the Chamaeleon~II cloud. Symbols as in
		Fig.~\ref{fig:crafield}.}
              \label{fig:chaiifield}
    \end{figure}
%

%
%########################################################################
\section{Observations}\label{sec:wfi}

%__________________________________________________  table
   \begin{table}
   \centering
   \footnotesize
      \caption[]{\footnotesize Log of WFI observations in Corona Australis and
      Chamaeleon~II}
         \label{tab:cralog}
     %$$ 
     \vspace{0.5cm}      
         \begin{tabular}{p{0.2\linewidth}l c c c}
            \hline
Date & Field & $\alpha$ (2000) & $\delta$ (2000) \\
            \hline
            \hline
28 May 1999 & CrA-2                & 19$^h$~00$^m$~04.0$^s$ & -37$^{\circ}$ 29$^{\prime}$ 34$\farcs$8 \\
28 May 1999 & CrA-4$^{\mathrm{a}}$ & 19$^h$~01$^m$~45.9$^s$ & -36$^{\circ}$ 58$^{\prime}$ 30$\farcs$0 \\
\hline
30 May 1999 & ChaII-3		   & 12$^h$~58$^m$~14.5$^s$ & -76\degr~50\arcmin~58$\farcs$7 \\
\hline
\end{tabular}
     %$$       
\footnotesize     
\begin{list}{}{}
\item[$^{\mathrm{a}}$] This field contains the Coronet cluster.
\end{list}
   \end{table}

%__________________________________________________  table
   \begin{table}
   \centering
   \footnotesize
      \caption[]{\footnotesize Exposure times (in seconds) for the images taken 
      		in each selected filter.}
         \label{tab:times}
     $$ 
         \begin{array}{p{0.2\linewidth}c c c c}
            \hline
Filter & $Short~exp.$ \ & $Interm.~exp.$ \ & $Long~exp.$ \ \\
            \hline
Rc/162 &  5 & 60 & 600 \\
Ic/lwp &  5 & 30 & 600 \\
Halpha/7  & 15 & $-$ & 600 \\
855/20 &  16 & 300 & 600 \\
915/28 &  8 & 100 & 600 \\
            \hline
         \end{array}
$$
   \end{table}

	Our optical survey was carried out in May 1999 using the Wide Field
Imager (WFI) at the ESO/MPG 2.2-m telescope on La Silla Observatory (Chile).
This eight-chip mosaic camera has a field of view of about
34\arcmin$\times$34\arcmin \ and a spatial resolution of 0.238\arcsec/pixel. In
Corona Australis, two WFI fields were observed, covering a total area of about
0.6~deg$^2$ in this dark cloud (see Fig.~\ref{fig:crafield}). One of the fields
(CrA-4) contained the intermediate-mass star R~CrA and the Coronet cluster. In
Chamaeleon~II, we observed only one WFI field, located to the north-west of the
densest part of the cloud (see  Fig.~\ref{fig:chaiifield}). Part of a dense
cloud core can be seen towards the western part of the field, which contains a
few previously known T~Tauri stars (Schwartz \cite{schwartz91}). 

	We observed in two broad-band filters, $R$ and $I$, in a narrow-band
filter centred on the H$\alpha$ emission line (H$\alpha$/7), and in two
medium-band filters, $M855$ (855/20) and $M915$ (915/28). The last two allow
for a photometric spectral type classification (see Sect.~\ref{sec:spt} below).
For each field and filter, except for H$\alpha$, we took three exposures with
different exposure times, in order to prevent the brightest T~Tauri stars (TTS)
from saturating. For the H$\alpha$ filter only two different exposure times
were used (see Table~\ref{tab:times}). The average seeing was $\sim 1\arcsec$.

	Data reduction, object search, and photometry were performed within the
IRAF environment.\footnote{\footnotesize IRAF is distributed by the National
Optical Astronomy Observatory (NOAO), operated by the Association of
Universities for  Research in Astronomy, Inc. under contract to the US National
Science Foundation. } The images were first bias-subtracted and flatfielded.
After this process, an irregular illumination pattern remained, which was
corrected by division through an illumination mask. The reddest filters ($I$,
$M855$, and $M915$) also showed a strong fringing pattern. To correct for it,
it was necessary to subtract a fringe mask from the reduced images. Both masks
were created by combination of the science exposures (see Scholz \& Eisl\"offel
\cite{scholz04} for more details). 

	An object catalogue of each surveyed field was produced by running 
SExtractor (Bertin \& Arnouts \cite{bertin96}) on the longest $I$ exposures. 
Aperture and PSF photometry was performed using standard routines from the 
DAOPHOT package (Stetson \cite{stetson87}). In this way we could also measure
the faintest components of multiple systems. For the short exposures and the
standard stars, we did only aperture photometry.

	The broad-band photometry was calibrated with Landolt
(\cite{landolt92}) standard stars. Our survey is complete down to
$R\simeq$20~mag and $I\simeq$19~mag. Since no set of standard stars is
available for our narrow- and medium-band filters, no calibration was possible
in these bands. A correction for atmospheric extinction was performed using the
extinction coefficients from the fit of the Landolt stars for the $R$-band, in
the case of the H$\alpha$ filter, and the $I$-band, for the $M855$ and $M915$
photometry. For a better understanding of the emission properties, the
H$\alpha$ instrumental magnitudes were then shifted so that the  main locus of
the objects in each surveyed field corresponds to H$\alpha-R=0$ in a
(H$\alpha$, H$\alpha-R$) colour-magnitude diagram (see Sect.~\ref{sec:cand}). 

	 Since only a global fit for all the CCD chips could be performed with
the Landolt standard stars, the main source of error in our photometric
calibration is produced by the WFI chip-to-chip photometric variations, which
can be as high as 3\% in the broad-band filters and 5\% in medium-band filters 
(Alcal\'a et al. \cite{alcala02}). On the other hand, as reported in
\cite{lm04}, a systematic offset exists between images with different exposure
times, in the sense of the resulting luminosities being brighter (typically by
about 0.05~mag) for longer exposure times. The origin of this offset is so far 
unknown. To minimize this systematic error, we determined the offsets between 
the three different integration times, and then shifted the short and long 
integration times to the system of the medium integration. 

	The listed magnitudes are the average of the so-corrected measurements
only if the average error was lower than the errors of the individual
measurements. If this was not the case, the measurement with the lower error
was preferred. We estimate that our survey is complete down to $R\simeq$21~mag
and $I\simeq$20~mag. The errors in the completeness range are in general not
larger than 5\%. 

	A more detailed discussion of the reduction, photometry and calibration
procedures and of the sources of the photometric errors can be found in
\cite{lm04}. 

%
%
%########################################################################
\section{Identification of new cloud members}\label{sec:bd}

%
%_____________________________________________________________
   \begin{figure}[t]
   \centering
   \includegraphics[width=6.5cm, angle=-90, bb= 50 70 600 775]{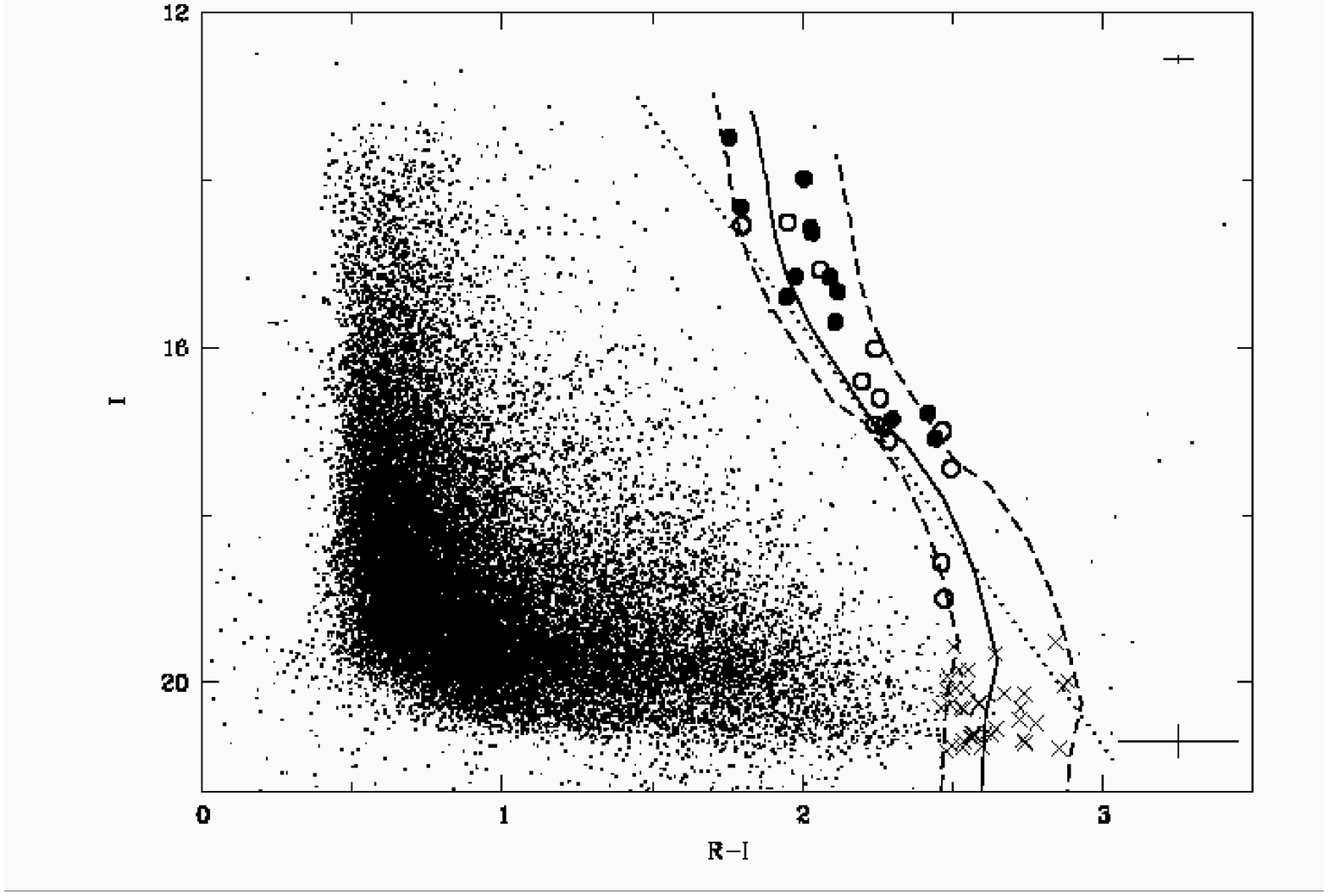}\hfill  
   \includegraphics[width=6.5cm, angle=-90, bb= 50 70 600 775]{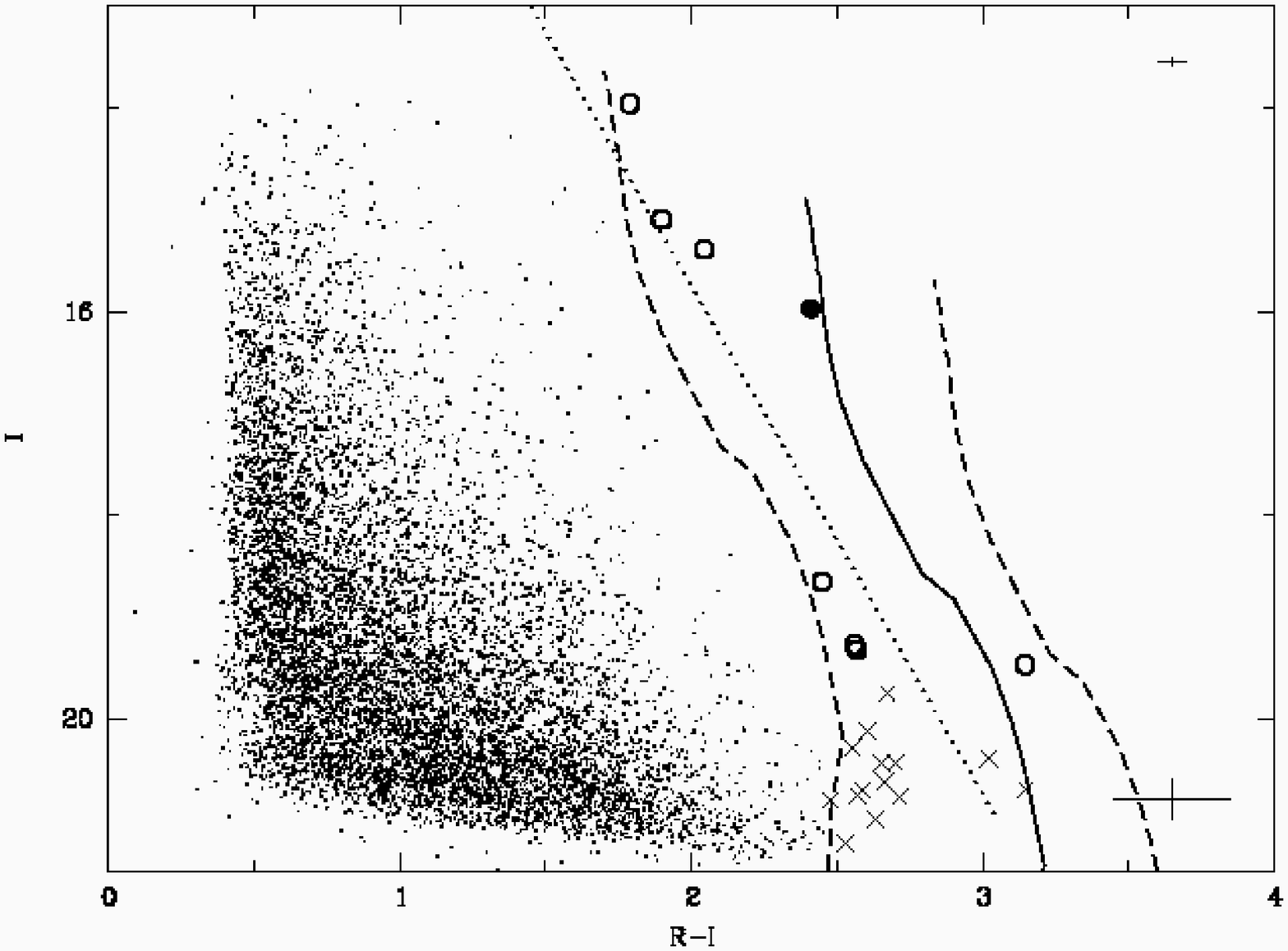}\hfill
      \caption{\footnotesize
	($I, R-I$) colour-magnitude diagrams for our surveyed fields in Corona
	Australis (upper panel) and Chamaeleon~II (lower panel). For clarity,
	only a random selection of the background objects has been plotted.
	Brown dwarf candidates are selected around a 1~Myr isochrone from the
	models of Chabrier et al. (\cite{chabrier00}) (solid line). Objects
	with and without detected H$\alpha$ emission are marked with solid and
	open circles, respectively. The objects marked with 'x' symbols are
	rejected (see text).The dashed lines indicate the range in colour
	around the isochrones taken as limit for the candidate selection, while
	the dotted line is the empirical isochrone used for candidate selection
	in Chamaeleon~I. The crosses indicate the average errors. The
	theoretical models of Chabrier et al. place the stellar/substellar
	boundary at I$\simeq$14-15~mag for the distances, extinctions, and ages
	considered.} 
	\label{fig:ridiag}
   \end{figure}
%

%______________________________________________________________
\subsection{Candidate selection} \label{sec:cand}

	To identify low-mass members of our surveyed clouds, we proceeded as in
our previous works (\cite{lm04} and \cite{lm05}). Candidates were selected in
an ($I, R-I$) colour-magnitude diagram around the position of a 1~Myr
theoretical isochrone from the Chabrier et al. (\cite{chabrier00}) models,
shifted to the estimated distance and extinction of the cloud.

	Average extinction values of
(A$_{\mathrm{V}})_{\mathrm{CrA}}$=0.47$^{+1.03}_{-0.47}$  and
(A$_{\mathrm{V}})_{\mathrm{ChaII}}$=2.2$^{+2.2}_{-2.2}$ were estimated using
the individual values measured by Neuh\"auser et al. (\cite{neuhauser00}) and
Vuong et al. (\cite{vuong01}) for two samples of stars seen in the direction of
Corona Australis and Chamaeleon~II, respectively. With the following relations:

\begin{center}
\begin{equation}\label{eq:ext1}
\mathrm{A}_{\mathrm{I}}/\mathrm{A}_{\mathrm{V}}=0.482
\end{equation}
\begin{equation}\label{eq:ext2}
\mathrm{A}_{\mathrm{I}}=1.812 \cdot [(\mathrm{R-I})-(\mathrm{R-I})_0],
\end{equation}
\end{center}

\noindent
derived from the extinction law of Rieke \& Lebofsky (\cite{rieke85}), average
colour excesses of

\begin{equation}\label{eq:colxs1}
\mathrm{E(R-I)_{CrA}}=0.13^{+0.28}_{-0.13} 
\end{equation}
\begin{equation}\label{eq:colxs2}
\mathrm{E(R-I)_{ChaII}}=0.69^{+0.44}_{-0.69} 
\end{equation}

\noindent
were computed for each region. 

	For the candidate selection, we took a 1~Myr theoretical isochrone from
the dusty models of Chabrier et al. (\cite{chabrier00}), shifted according to
the distance and our estimated average extinction for each dark cloud.
Fig.~\ref{fig:ridiag} shows the ($I, R-I$) colour-magnitude diagrams for our
surveyed fields. Our brown dwarf candidates are the objects found around each
reddened isochrone, with a maximum scatter in colour given by the lower and
upper limit of the average colour excess as determined above. All objects found
within the given colour ranges were selected as candidate low-mass members of
the clouds. For comparison, the diagrams in Fig.~\ref{fig:ridiag} also show the
position of the empirical isochrone used for candidate selection in
Chamaeleon~I (dotted  line). Note that this line is placed towards bluer (R--I)
colours than the theoretical Chamaeleon~II isochrone (see lower panel in
Fig.~\ref{fig:ridiag}). This is consistent with the lower distance (160~pc) and
extinction of Chamaeleon~I with respect to Chamaeleon~II. 

	According to the theoretical models by Chabrier et al.
(\cite{chabrier00}), if these objects indeed belong to Corona Australis and
Chamaeleon~II, they must have masses ranging from the hydrogen burning limit
($\sim 0.075$-0.080~M$_{\odot}$) to about 0.005~M$_{\odot}$. However, the large
photometry errors at the lower part of these diagrams introduce some
uncertainty in the membership of the faintest objects: As seen in
Fig.~\ref{fig:ridiag}, a large number of the selected objects in both clouds
have $I>$19.5~mag. With the assumed values of distance and extinction, the
models give masses for them of M$<$0.012M$_{\odot}$ in Chamaeleon~II and
M$<$0.007M$_{\odot}$ in Corona Australis, which would place them below the
deuterium burning mass limit ($\sim$0.013M$_{\odot}$). However, most of these
very faint candidates are found near the edge of our selection band; thus, a
slight shift towards bluer colours caused by the photometric errors would place
some of these objects outside our selection area in the colour-magnitude
diagram. Moreover, these objects are also very faint in H$\alpha$, or not
detected at all in this band. Taken all this into account, we conclude that
they are more likely background objects, and remove them from our candidate
lists for Corona Australis and Chamaeleon~II. These objects are marked with 'x'
in the diagrams of Fig.~\ref{fig:ridiag}, and won't be further considered. 

	In principle, a high contamination from older objects is not expected
in the direction of these star forming regions. Their  proximity and high
Galactic latitude ($b\approx-17^{\circ}$) imply that the density of foreground
objects must be very low in both cases. Moreover, the extinction suppresses
detection of objects behind the dark clouds. Particularly, background
contamination should be negligible towards the core seen to the north-west of
our Chamaeleon~II field (Fig.~\ref{fig:chaiifield}), because the extinction is
larger in this area (see e.g. Vuong et al. \cite{vuong01}; Mizuno et al.
\cite{mizuno99}). This is consistent with the low number of objects detected in
the area of interest of the corresponding ($I, R-I$) colour-magnitude diagram
(lower panel in Fig.~\ref{fig:ridiag}), once the faintest objects are discarded
for the reasons explained above.

%
%_____________________________________________________________
  \begin{figure}[t]
   \centering
   \includegraphics[width=6.5cm, angle=-90, bb= 50 70 600 775]{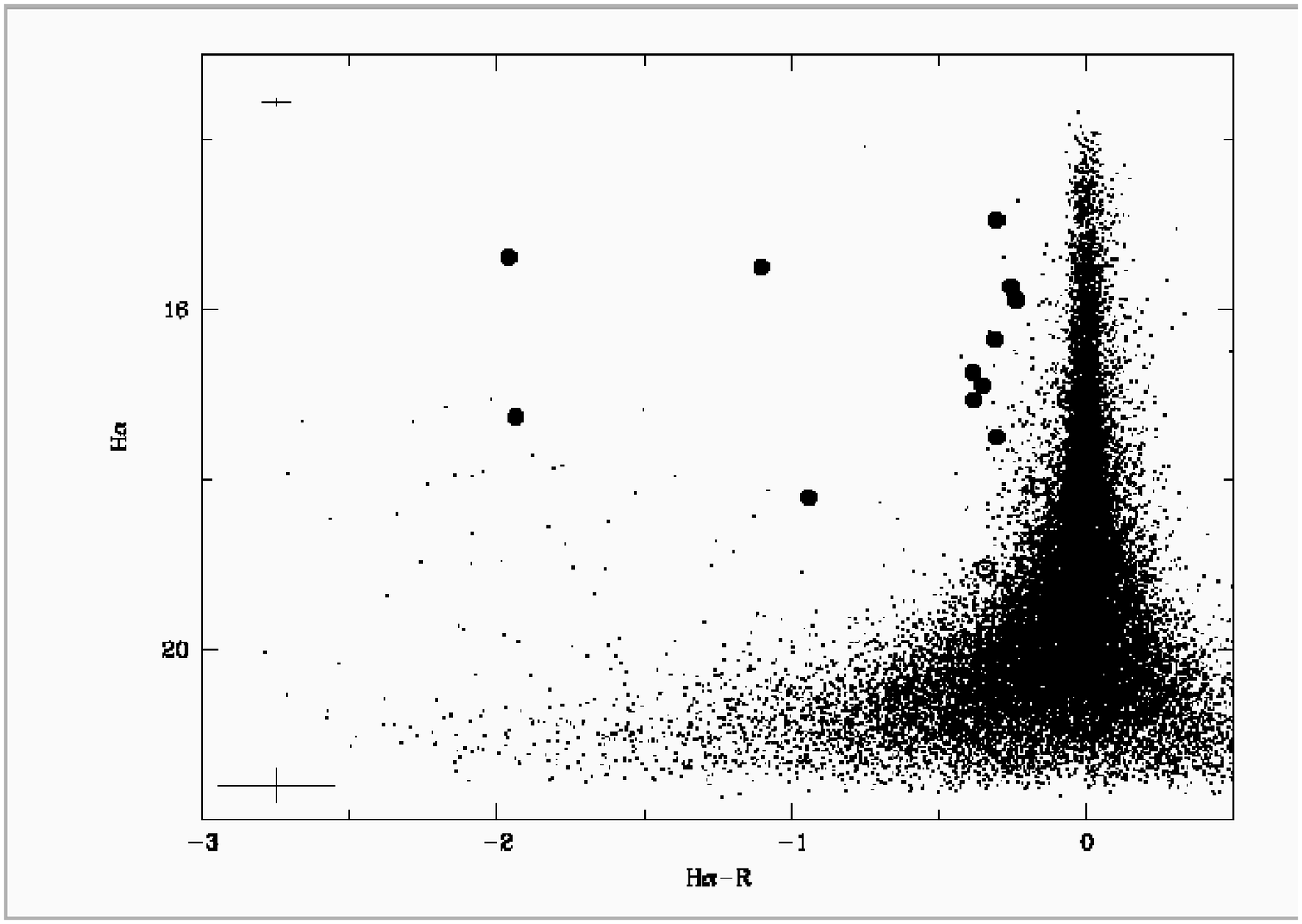}\hfill
   \includegraphics[width=6.5cm, angle=-90, bb= 50 70 600 775]{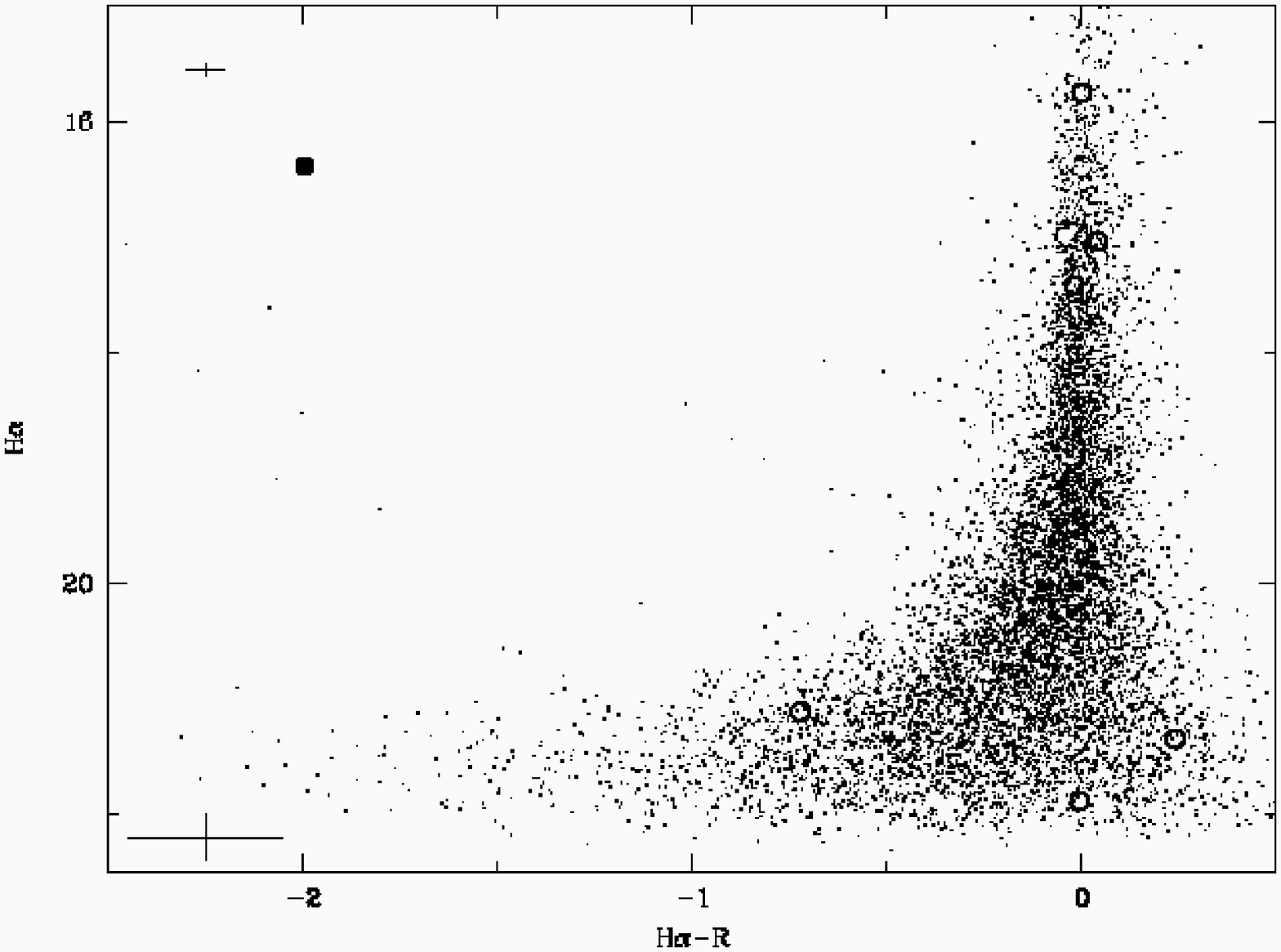}\hfill
      \caption{\footnotesize
	(H$\alpha$, H$\alpha-R$) colour-magnitude diagrams for our surveyed
	fields in Corona Australis (upper panel) and Chamaeleon~II (lower
	panel). Symbols as in Fig.~\ref{fig:ridiag}.} 
      \label{fig:hardiag}
   \end{figure}
%

%______________________________________________________________
\subsection{Probable members} \label{sec:mem}

	In a second step, probable cloud members were selected by means of
their H$\alpha-R$ colour. As shown in \cite{lm04}, this colour is a useful
estimation of the level of H$\alpha$ emission, a common youth diagnostics.
Hence, if our candidates showed H$\alpha$ emission, they were likely to be
young objects, and thus members of the star-forming region.

	Fig.~\ref{fig:hardiag} shows the (H$\alpha$, H$\alpha-R$)
colour-magnitude diagrams for our surveyed areas in Corona Australis and
Chamaeleon~II. For the brightest objects, we applied the same criterion as in
previous works (Lamm et al. \cite{lamm05}; \cite{lm05}), retaining the objects
with

\begin{equation}\label{halcrit}
\Delta(\mathrm{H}\alpha-\mathrm{R})=
(\mathrm{H}\alpha-\mathrm{R})_{\mathrm{object}}
-(\mathrm{H}\alpha-\mathrm{R})_{\mathrm{locus}}
<-0.1 
\end{equation}

\noindent
as probable cloud members.

	In the case of Chamaeleon~II, this procedure left us with only one good
candidate, ChaII~376. This object is found to have clear H$\alpha$ emission,
its H$\alpha-R$ colour being as negative as those exhibited by some brown
dwarfs and very low-mass stars in Chamaeleon~I (see Fig.~3 in \cite{lm04}).
However, as outlined below (Sect.~\ref{sec:spt}), its derived spectral type is
inconsistent with a brown dwarf. In Corona Australis, 13 candidates are
retained as probable cloud members.

	The non-detection of H$\alpha$ emission does not completely rule out
the rest of the objects as members of Corona Australis and Chamaeleon~II. For
instance, in the slightly older (2-4~Myr) open cluster NGC~2264, Lamm et al.
(\cite{lamm05}) found a considerable number of very low-mass stars on the basis
of their variability, many of which showed no detectable H$\alpha$ emission.
Indeed, we later added one object to the Corona Australis and two to the
Chamaeleon~II candidate lists based on their infrared properties (see
Sect.~\ref{sec:nir} below). The rest need confirmation of their status by
follow-up observations.

%
%
%_____________________________________________________________
   \begin{figure}[t]
   \centering
  \includegraphics[width=6.5cm, angle=-90, bb= 50 70 600 775]{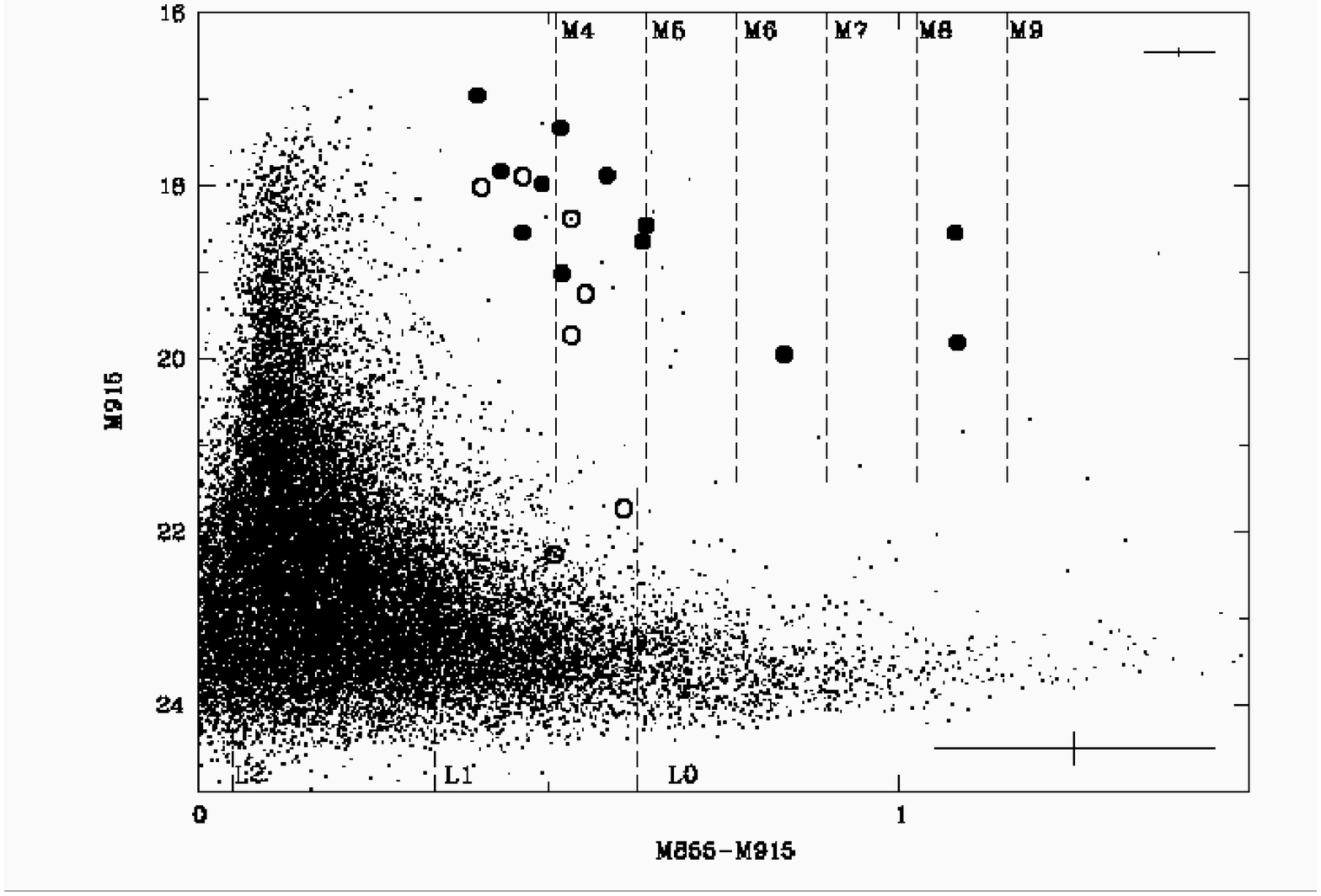}\hfill
  \includegraphics[width=6.5cm, angle=-90, bb= 50 70 600 775]{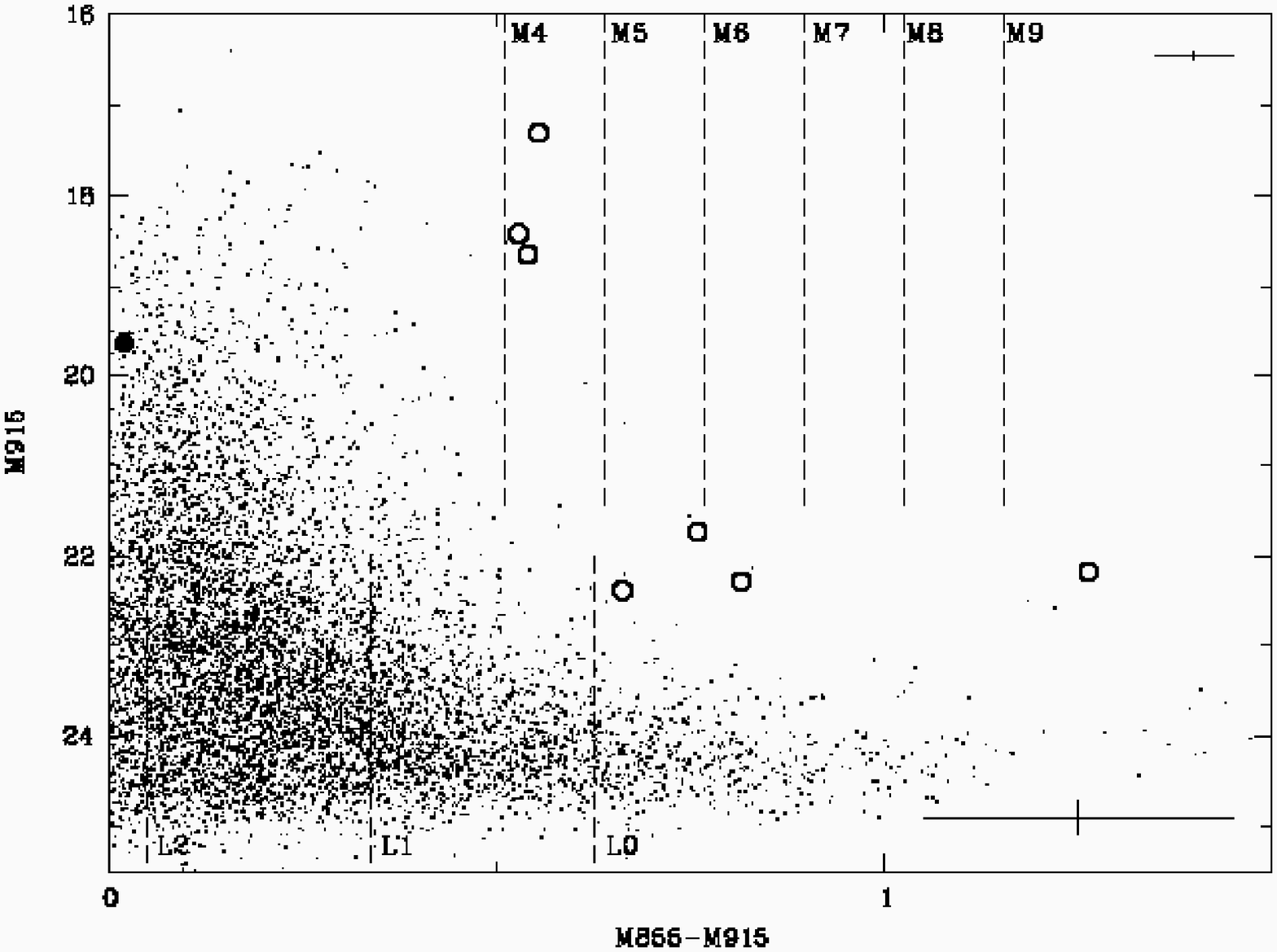}\hfill
      \caption{\footnotesize
	($M915$, $M855-M915$) colour-magnitude diagrams for our surveyed fields
	in Corona Australis (upper panel) and Chamaeleon~II (lower panel).
	Symbols as in Fig.~\ref{fig:ridiag}. Our scale for the identification
	of the spectral type is also indicated. The faintest objects in the
	diagrams are regarded as possible L-type objects.}
         \label{fig:mmdiag}
   \end{figure}
%
%

%______________________________________________________________
\subsection{Spectral types}\label{sec:spt}

	To identify the spectral types of our candidates, we followed the
method outlined in \cite{lm04}, based on the observations of field M- and early
L-dwarfs in the two medium-band filters $M855$ and $M915$. The former lies in a
spectral region including the TiO and VO absorption bands used to classify
medium and late M-type objects (see Kirkpatrick et al. \cite{kirkpatrick91}).
The filter $M915$ covers a pseudo-continuum region in these late-type objects.
In \cite{lm04}, a correlation was found between the $M855$--$M915$ colour index
and the spectral type for the range M4-M9. This enabled us to derive spectral
types for our brown dwarf candidates from their position in a ($M915$,
$M855-M915$) colour-magnitude diagram. We estimated an error of 1-2 subclasses
in this calibration. For spectral types earlier than M4 the $M855$--$M915$
colour saturates, because the TiO and VO features are not prominent or not
present at all in the spectra. Therefore, it is not possible to determine
spectral types earlier than M4 with this method.

	A second correlation, of opposite sign, was found for spectral types
L0-L2, due to the progressive disappearance of the TiO and VO from the spectra.
In this case, however, the uncertainties are larger than for late\,M-type
objects, given that extincted early\,M-objects would be placed in the same
region of the colour-magnitude diagram. 

	Based on comparison with published spectral types for some previously
identified Chamaeleon~I members, in \cite{lm04} we estimated that about 80\% of
the spectral types derived with this photometric method are correct (within the
quoted errors). Wrong spectral types would be derived, though, for highly
extincted objects and unresolved binaries. Comer\'on et al. (\cite{comeron04})
report an objective prism survey of the entire Chamaeleon~I cloud, providing
long-slit spectroscopy and $VRIJHKs$ imaging of the identified candidates.
Seven of their objects are also included in our Chamaeleon~I survey, and our
given spectral types agree fairly well with the ones provided by these authors.
Luhman (\cite{luhman04}) provides optical spectroscopy for 179 possible members
of this dark cloud. Of the 25 objects in common with our own survey, 17 have
spectral types that are in good agreement with our photometric classification.
For another four objects, Luhman (\cite{luhman04}) gives spectral types which
are still marginally coincident with our predictions. Hence, 100\% of our
objects in common with Comer\'on et al. (\cite{comeron04}) and at least about
68\% of our objects in common with Luhman (\cite{luhman04}) have
photometrically derived spectral types confirmed and refined by these authors.
This makes a total of 75\% of coincidence between the spectroscopic and the
photometric spectral types (within the estimated errors), in fairly good
agreement with the expectations. The discrepancies correspond to earlier type
objects for which our method yielded a much later spectral type, probably due
to local high extinction towards them.

A detailed discussion of this photometric spectral type classification, its
error sources and its limitations can be found in \cite{lm04}.

%__________________________________________________  table
\begin{table*}[t]  
\centering
\scriptsize
     \caption[]{\footnotesize Candidate low-mass members of 
       Corona Australis and Chamaeleon~II$^{\mathrm{a}}$$^{\mathrm{b}}$}
         \label{tab:wfiphot}
      \vspace{0.3cm}

         \begin{tabular}{p{0.1\linewidth}c c c c c c c l l l}
            \hline
Name & $\alpha$ (2000) & $\delta$ (2000) & $R$ & $I$ & H$\alpha$ & $M855$ & $M915$ & SpT$^{\mathrm{c}}$ & Classification & Other names$^{\mathrm{d}}$ \\
 & $hh~mm~ss.s$ & $ddd~mm~ss.s$ & mag & mag & mag & mag & mag & & & \\
            \hline
            \hline
\object{CrA~205}   & 19 01 11.5  & -37 22 21.1 &  17.12 &  15.15 &  16.74 &  19.00 & 18.54 & M4 & star & \\
\object{CrA~432}   & 19 00 59.8  & -36 47 09.2 &  19.16 &  16.86 &  18.21 &  20.78 & 19.95 & M7 & BD & \\
\object{CrA~444}   & 19 00 45.3  & -37 11 47.3 &  17.34 &  15.39 &  15.38 &  19.62 & 18.54 & M8.5 & BD cand. & \\
\object{CrA~452}   & 19 00 44.6  & -37 02 09.5 &  15.25 &  13.49 &  14.96 &  17.35 & 16.96 & $<$M4 & star & \object{ISO-CrA~98} \\
\object{CrA~453}   & 19 01 04.6  & -37 01 28.4 &  17.80 &  15.69 &  17.49 &  19.53 & 19.01 & M4 & star    & \\
\object{CrA~465}   & 19 01 53.7  & -37 00 33.6 &  19.20 &  16.78 &  17.26 &  20.90 & 19.81 & M8.5 & BD cand.& \object{WMB~185831.1-370456} \\
\object{CrA~466}$^{\mathrm{e}}$ & 19 01 19.0  & -36 58 27.2 &  18.25 &  16.01 &  18.09 &  19.80 & 19.25 & 4.5 & star & \\
\object{CrA~468}   & 19 01 49.3  & -37 00 28.1 &  16.12 &  14.33 &  15.88 &  18.26 & 17.83 & $<$M4 & star   & \object{WMB~185826.8-370450} \\
\object{CrA 4107}  & 19 02 54.6  & -36 46 18.9 &  16.60 &  14.57 &  15.50 & 
18.46 & 17.88 & M4.5 & star & \object{ISO-CrA~177} \\
\object{CrA 4108}  & 19 02 09.7  & -36 46 33.5 &  16.66 &  14.63 &  16.35 & 
18.46 & 17.97 & M4 & star & \object{ISO-CrA~141} \\
\object{CrA 4109}  & 19 02 16.7  & -36 45 48.5 &  15.99 &  13.98 &  15.73 & 
17.85 & 17.33 & M4 & star & \object{ISO-CrA~146} \\
\object{CrA 4110}  & 19 01 16.3  & -36 56 27.3 &  17.25 &  15.16 &  16.89 & 
19.09 & 18.45 & M5 & star & \object{ISO-CrA~123} \\
\object{CrA 4111}  & 19 01 20.8  & -37 03 02.3 &  17.45 &  15.33 &  17.06 & 
19.28 & 18.64 & M5 & star & \\
\hline 
\object{ChaII~376} & 13 03 13.3 & -76 50 50.9 & 18.38 & 15.97 & 16.39 & 19.66 & 19.64 & $<$M4 & star & \\
\object{ChaII~304}$^{\mathrm{e}}$ & 12 55 15.9 & -76 46 21.8 & 21.8; & 19.27 & 21.1; & 23.1; & 22.3; & M9.5 & BD cand. & \\
\object{ChaII~305}$^{\mathrm{e}}$ & 12 55 16.3 & -76 46 20.9 & 21.9; & 19.33 & 21.9; & 23.0; & 22.4; & M9.5 & BD cand. & \\
\hline
\end{tabular}
     %$$     
\begin{list}{}{}
\item[$^{\mathrm{a}}$] H$\alpha$, $M855$, and $M915$ magnitudes are
instrumental  magnitudes.
\item[$^{\mathrm{b}}$] Photometric errors: blank: 0.05~mag; semicolon: 0.1~mag;
colon: 0.2~mag
\item[$^{\mathrm{c}}$] Errors in the spectral types: M4-M6: 2 subclasses; 
M6.5-M9: 1 subclass
\item[$^{\mathrm{d}}$] References: WMB\#: Wilking et al.(\cite{wilking97});
	 ISO-CrA\#: Olofsson et al. (\cite{olofsson99})
\item[$^{\mathrm{e}}$] Infrared selected (see text).
\end{list}
   \end{table*}

	Fig.~\ref{fig:mmdiag} shows the ($M915$, $M855-M915$) colour-magnitude
diagrams for our surveyed fields in Corona Australis and Chamaeleon~II. Many of
our brightest candidates in both clouds have spectral types earlier than M6
according to these diagrams. Hence, if they belong to the cloud, they must be
low-mass stars. Only a few H$\alpha$ emitters in Corona Australis with spectral
types between M6.5 and M9 could be brown dwarfs. Some non-emitters in both
clouds could also have spectral types corresponding to the brown dwarf regime.

	The $M855$--$M915$ colour of our single clear H$\alpha$ emitter in
Chamaeleon ~II, ChaII~376, is remarkably bluer than expected for M-type objects
($M855-M915=-0.02$). This might be an indication of a very early spectral type.
The object might be either an extincted early-type member of Chamaeleon~II or
an older H$\alpha$ emitting object, such as an AGB star. Even if its membership
of the star forming region were confirmed, ChaII~376 is not likely to be a
brown dwarf. 	

	The WFI photometry and the estimated spectral types for all the 
probable members of Corona Australis (as well as an infrared selected
candidate, see Sect.\ref{sec:nircra} below) are summarized in
Table~\ref{tab:wfiphot}. This table also includes the WFI photometry for
ChaII~376 and other possible members of Chamaeleon~II (see
Sect.~\ref{sec:nirchaii}). Note that only  three objects in Corona Australis
and two in Chamaeleon~II have estimated spectral types that are clearly
substellar for the age of these regions.

%______________________________________________________________
\subsection{Infrared detections}\label{sec:nir}

	We made use of the database of the 2MASS survey\footnote{\footnotesize
Available online at the URL of the NASA-IPAC Infrared Science Archive (IRSA):
{\tt http://irsa.ipac.caltech.edu/}}  and of published results from other
authors to look for near-infrared counterparts of the objects (with and without
H$\alpha$ emission) in our initial candidate lists for both clouds. Moreover,
we also checked whether some  of our optical sources had been detected in the
mid-infrared by ISOCAM or Spitzer. In this way, we intended not only to further
confirm the youth of our H$\alpha$ emitters, but also to identify new
candidate cloud members.

%-------------------------------------------------------------------------
\subsubsection{Infrared sources in Corona Australis}\label{sec:nircra}

	All the probable Corona Australis members from Table~\ref{tab:wfiphot}
have a near-infrared counterpart within about 1$\arcsec$, except CrA~205,
CrA~432, and CrA~452, for which the closest 2MASS source is found at about 1.9,
2.1, and 1.5$\arcsec$, respectively. However, all three objects are placed in
areas of high cloud density, and no other optical source is seen in our WFI
images that could correspond to the near-infrared detection. We therefore
identified the 2MASS sources with our new cloud members. The near-infrared
photometry for all our H$\alpha$ emitters (and one object with mid-infrared
excess, see below) is summarized in Table~\ref{tab:irphot}. Two of these
objects (CrA~465 and CrA~468) had also previous near-infrared photometry
available from Wilking et al. (\cite{wilking97}). In addition, six objects
without H$\alpha$ emission from our initial candidate selection also have 2MASS
counterparts.

\begin{figure}[t]
   \centering
  \includegraphics[width=6.cm, angle=-90, bb= 50 70 600 775]{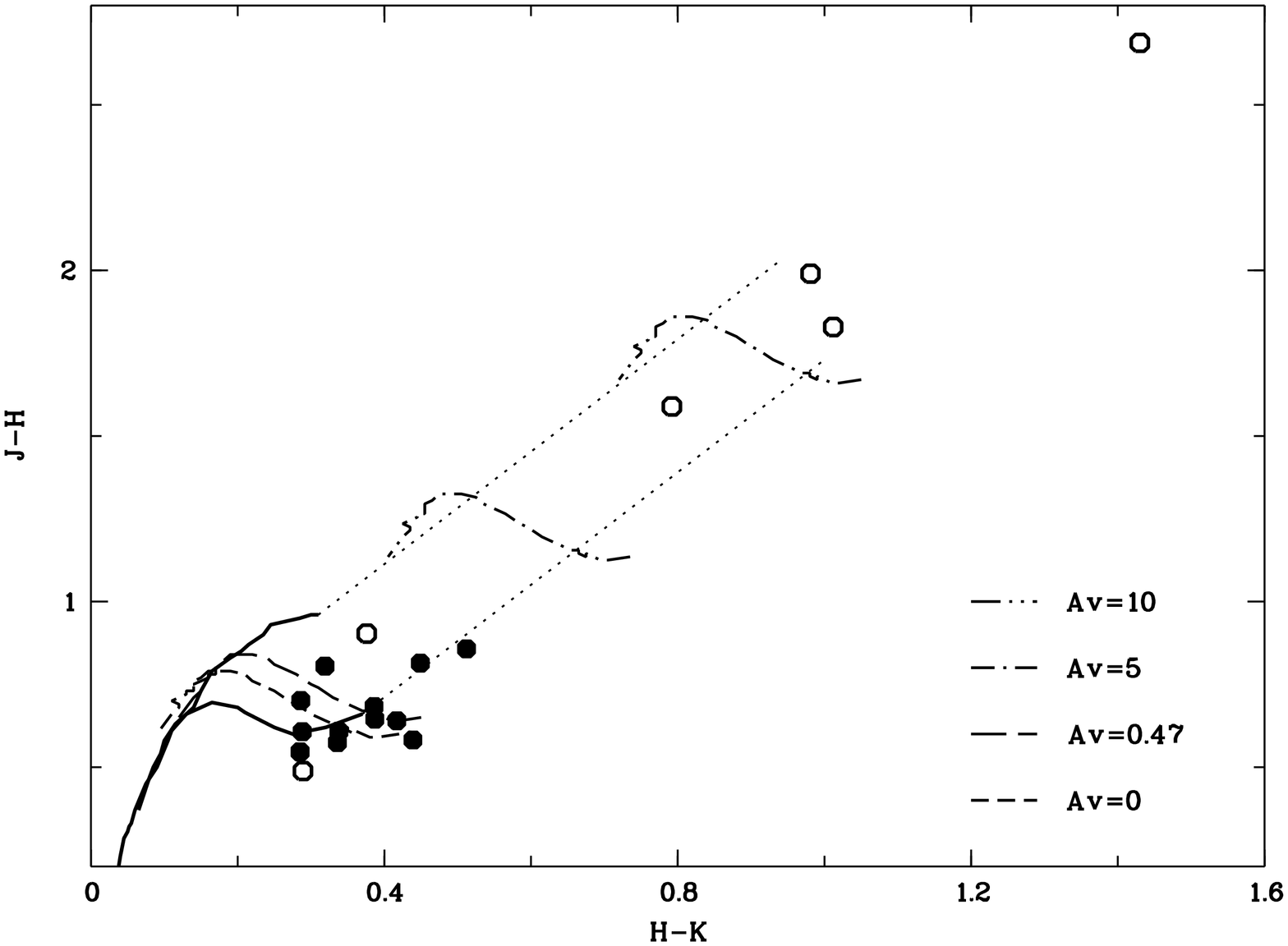}\hfill
  \includegraphics[width=6.cm, angle=-90, bb= 50 70 600 775]{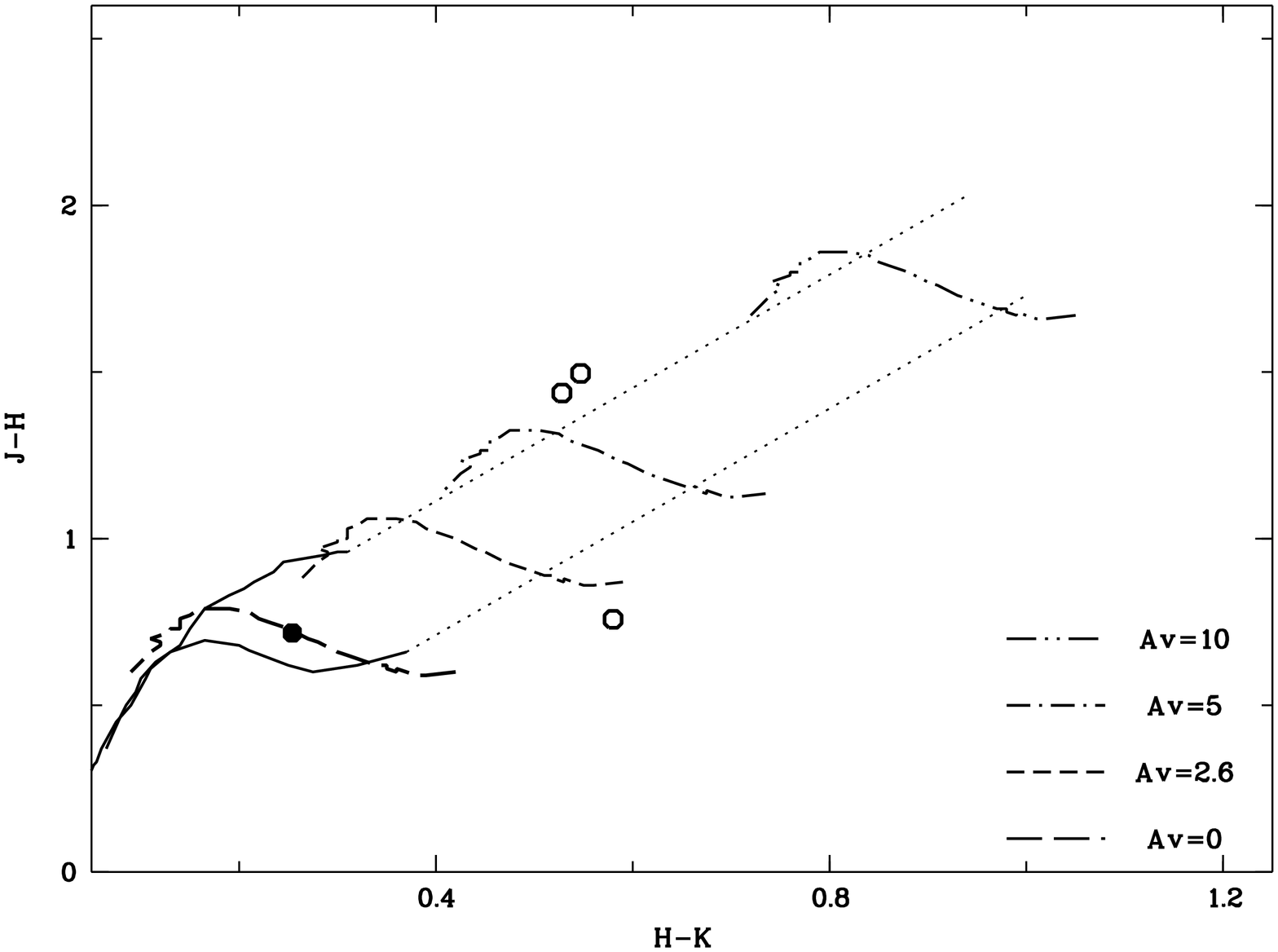}\hfill
      \caption{\footnotesize
	Colour-colour diagrams for our candidates in Corona Australis (upper
	panel) and Chamaeleon~II (lower panel) with near-infrared  photometry.
	Objects with and without H$\alpha$ emission are marked with solid and
	open symbols, respectively. The solid curves indicate the locus of
	dereddened dwarfs and giants according to Bessell \& Brett 
	(\cite{bessell88}). The dotted lines indicate the direction of the
	reddening vector up to A$_\mathrm{V}\sim$10~mag. The other lines
	indicate the position of a 1~Myr isochrone from the models of Baraffe
	et al. (\cite{baraffe98}) for different extinction values.
      }
         \label{fig:nirdiag}
   \end{figure}

	The upper panel of Fig.~\ref{fig:nirdiag} shows the $(J-H, H-K)$
colour-colour diagram for all the 2MASS detected objects in Corona Australis.
The locus of dereddened dwarfs and giants (Bessell \& Brett \cite{bessell88})
is also indicated, as well as the direction of the reddening vector and the
positions of a 1~Myr isochrone from the models of Baraffe et al.
(\cite{baraffe98}) for different values of extinction. We see that most of the
objects without H$\alpha$ emission are placed at high values of extinction
(A$_V \gtrsim$ 10~mag), while the H$\alpha$ emitters all have low extinction,
consistent with our selection criterion in Sect.~\ref{sec:cand}. However, since
none of our objects has a near-infrared excess, we cannot confirm their young
status. 

	Among the H$\alpha$ emitters, the most extincted objects are CrA~432
and CrA~453 (A$_V\simeq$1.5~mag). The former has an estimated spectral type of
M7 and is considered a brown dwarf candidate in Table~\ref{tab:wfiphot};
however, according to the models of Chabrier et al. (\cite{chabrier00}), our
estimated mean cloud extinction, and the temperature scale of Luhman
(\cite{luhman99}), and with an extinction of  A$_V=$1.5~mag, CrA~432 should
have a spectral type of about M8.5, even later than our classification.  It
might be that this apparent underluminosity is intrinsic to the object itself,
as in the case of another member of the R~CrA cloud, LS-RCrA~1 (Fern\'andez \&
Comer\'on \cite{fernandez01}). Possible causes could be extinction by
surrounding dust or accretion processes. Also CrA~453 seems too faint for its
estimated spectral type (M4).

	Corona Australis has also been observed in the mid-infrared by the ISO 
satellite (Olofsson et al. \cite{olofsson99}). Unfortunately, the ISOCAM 
observations could not map the densest part of the cloud core, because the
brightest stars would have saturated the detector. Therefore, the overlap with
our own survey is small. Still, we find five objects in common with these
authors: CrA~452 (ISO-CrA~98), CrA~466 (ISO-CrA~127), CrA~4107 (ISO-CrA~177),
CrA~4109 (ISO-CrA~146) and CrA~4110 (ISO-CrA~123). The mid-infrared photometry
for all these objects is listed in Table~\ref{tab:irphot}. 

	With a single exception, all the ISOCAM detected sources have H$\alpha$
emission according to their H$\alpha-R$ colour. The position of the only
non-emitter, CrA~466, in the near-infrared colour-colour diagram could
correspond to a highly extincted young star or to a field dwarf. In the WFI
mosaic, this object is found at the upper right corner of the CCD\#55. No other
object is seen nearby, and the mismatch in the coordinates of the optical and
the mid-infrared source is of only 1$\arcsec$. However, given the gaps between
the different CCDs and the pointing errors of the ISOCAM positions, we cannot
exclude that the true counterpart of the ISOCAM source is another star not seen
in our images. Still, with a H$\alpha-R$ colour of --0.16, H$\alpha$ emission
from this object cannot be completely ruled out, given the errors in that part
of the colour-magnitude diagram (upper panel of Fig.~\ref{fig:hardiag}).
CrA~466 is further discussed in Sect.~\ref{sec:irpropcra}.

%
%_____________________________________________________________
%

%__________________________________________________  table
\begin{table*}[t]  
\centering
\scriptsize
     \caption[]{\footnotesize Infrared photometry for our candidate members of 
     Corona Australis and Chamaeleon~II$^{\mathrm{a}}$}
         \label{tab:irphot}
      \vspace{0.3cm}

         \begin{tabular}{p{0.1\linewidth}c c c c r r r}
            \hline
Name   & 2MASS Id. &   J   &  H  &   K   &  F$_{6.7}$  &  F$_{14.3}$ \\
  &  &   mag   &  mag  &   mag   &  mJy  &  mJy \\
            \hline
            \hline
CrA~4107      & 19025464-3646191 & 12.440 & 11.795 & 11.408 &  1.07 &  7.30 \\
CrA~4108      & 19020967-3646344 & 12.580 & 11.972 & 11.634 &       &       \\
CrA~4109      & 19021667-3645493 & 12.004 & 11.303 & 11.017 &  5.60 &       \\
CrA~4110      & 19011629-3656282 & 12.954 & 12.314 & 11.897 &  6.50 &       \\
CrA~4111      & 19012083-3703027 & 13.233 & 12.687 & 12.402 &       &       \\
CrA~205       & 19011169-3722213 & 13.315 & 12.741 & 12.405 &       &       \\
CrA~432       & 19005974-3647109 & 14.190 & 13.333 & 12.821 &       &       \\
CrA~444       & 19004530-3711480 & 12.924 & 12.342 & 11.903 &       &       \\
CrA~452       & 19004455-3702108 & 11.571 & 10.766 & 10.447 &  8.60 &       \\
CrA~453       & 19010460-3701292 & 13.338 & 12.524 & 12.075 &       &       \\
CrA~465$^{\mathrm{b}}$  & 19015374-3700339 & 14.084 & 13.401 & 13.015 &       &       \\
CrA~466       & 19011893-3658282 & 12.834 & 11.245 & 10.453 & 31.00 & 30.00 \\
CrA~468$^{\mathrm{c}}$  & 19014936-3700285 & 12.498 & 11.891 & 11.603 &       &       \\
\hline 
ChaII~304/305 & 12551617-7646213 & 15.950 & 15.192 & 14.612 &       &       \\
ChaII~376     & 13031246-7650509 & 14.337 & 13.620 & 13.366 &       &       \\
\hline
\end{tabular}
\begin{list}{}{}
\item[$^{\mathrm{a}}$] Near-infrared photometry from 2MASS; mid-infrared
photometry from ISOCAM (Olofsson et al. \cite{olofsson99}).
\item[$^{\mathrm{b}}$] Wilking et al. (\cite{wilking97}) report for this object:
$J=$14.06; $H=$13.44; $K=$13.09
\item[$^{\mathrm{c}}$] Wilking et al. (\cite{wilking97}) report for this object:
$J=$12.54; $H=$11.81; $K=$11.64
\end{list}
   \end{table*}

%-------------------------------------------------------------------------
\subsubsection{Infrared sources in Chamaeleon~II}\label{sec:nirchaii}

	Vuong et al. (\cite{vuong01}) report a list of objects in Chamaeleon~II
that have near-infrared excesses according to DENIS $IJKs$ photometry. We only
find two objects in common with these authors (\object{C30} and \object{C31}),
which represents 10\% of the DENIS sources in the overlapping area. If they
indeed belonged to Chamaeleon~II, both of these DENIS objects would have
spectral type M4 according to our calibration. Hence, they would be low-mass
stars. However, neither of them is a H$\alpha$ emitter according to our WFI
photometry.  

	We also searched for near-infrared counterparts to our Chamaeleon~II
objects in the  database of the 2MASS survey. We found five positive
detections, including ChaII~376 and both of our DENIS objects. 

	The lower panel of Fig.~\ref{fig:nirdiag} shows the ($J-H$, $H-K$)
colour-colour diagram for our 2MASS detected objects in Cha~II. The locus of
dereddened dwarfs and giants (Bessell \& Brett \cite{bessell88}) is also
indicated, as well as the direction of the reddening vector and the positions
of a 1~Myr isochrone from the models of Baraffe et al. (\cite{baraffe98}) for
different values of extinction. Most of our near-infrared detected candidates
have optical extinction values $A_V$ in the range 2-5~mag according to this
diagram. These values are in agreement with our selection criteria from
Sect.~\ref{sec:cand}. No evident excess emission can be inferred from the
near-infrared colours, as the objects are all placed along the reddening band.
Hence, we cannot separate cloud members and non-members with these data.

	The position of ChaII~376 in Fig.~\ref{fig:nirdiag} is consistent with
that of an unreddened $0.030\,M_{\odot}$ brown dwarf (M8) according to the
theoretical 1~Myr isochrone. However, its optical colours are bluer than
expected for this age and mass, and suggest a more massive substellar object
($M\sim0.060\,M_{\odot}$). Moreover, in Sect.~\ref{sec:spt} it was not possible
to assign a spectral type to this object by means of its ($M855-M915$) colour.
In view of these considerations, and since H$\alpha$ emission is not exclusive
of young objects, ChaII~376 might be an older active star. The 2MASS photometry
of this object is given in Table~\ref{tab:irphot}.

	Two objects, ChaII~304 and ChaII~305, have the same 2MASS counterpart.
They are seen close to each other in the WFI images ($\sim 2\arcsec$) and have
similar brightness and colours. Visual inspection confirmed that the pair is
not resolved in the 2MASS $J$-band image (which has a pixel size of
$2\arcsec$). Thus, the  near-infrared photometry provided in
Table~\ref{tab:irphot} seems to correspond to the unresolved pair.  In the
($J-H$, $H-K$) colour-colour diagram, the 2MASS counterpart is placed towards
the right of the reddening band at an extinction value of about 2.6~mag (the
average value used for candidate selection in Sect.~\ref{sec:bd}).
Interestingly, both members of this visual pair have similar optical
photometry, consistent with two objects close to or right below the deuterium
burning mass limit (0.015-0.012\,$M_{\odot}$) according to the Chabrier et al.
(\cite{chabrier00}) models. The near-infrared colours of the 2MASS detection,
however, are consistent with a single object of this mass. Given that no close
neighbour is seen in the 2MASS image, this may indicate that the unresolved
pair is not a very low-mass object in Chamaeleon~II. A further possibility is
that one of the two objects from our WFI images is a background star whose
near-infrared magnitudes are fainter than the 2MASS detection threshold. In
this latter case, the remaining object (most likely ChaII~304, whose position
is closer to the 2MASS source) would have optical \emph{and} near-infrared
colours consistent with a planetary mass object ($\sim 0.012\,$M$_{\odot}$) in
Chamaeleon~II (according to the Chabrier et al. models). The WFI photometry for
ChaII~304 and ChaII~305 is summarized in Table~\ref{tab:wfiphot}, as well as
the estimated spectral type and their classification if they were members of
the dark cloud.

	The two DENIS objects C30 and C31 are placed along the direction of the
reddening vector for giants. Both objects show a featureless optical spectrum
according to the observations by Barrado y Navascu\'es \& Jayawardhana
(\cite{barrado04b}), which did not allow these authors to conclusively confirm
or exclude their membership to Chamaeleon~II. Although these objects could
still be extincted cloud members (A$_{\mathrm{V}}\sim$5~mag), it seems more
probable, in view of their position in the ($J-H$, $H-K$) diagram, that they
belong to the background.\footnote{\footnotesize Barrado y Navascu\'es \&
Jayawardhana  (\cite{barrado04b}) compare the 2MASS photometry of the DENIS
sources provided by  Vuong et al. (\cite{vuong01}) with that of confirmed
Chamaeleon~II members, showing that, in general, the DENIS objects lie above
the main CTTS locus (Figure 4b in that work).}  
The other objects without H$\alpha$ emission have ambiguous positions in the
($J-H$, $H-K$) colour-colour diagram: They could be moderately extincted
Chamaeleon~II members or background dwarfs.

	We also cross-correlated our candidate list with that of the
Chamaeleon~II ISOCAM sources provided by Persi et al. (\cite{persi03}), but
none of our objects has a mid-infrared excess according to these authors. 
Young et al. (\cite{young05}) do not detect any of our candidates either in
their Spitzer-MIPS survey, covering more area at longer wavelengths and with
greater sensitivity.

%______________________________________________________________
\subsection{Object classification}\label{sec:class}

	After the selection process described above, we are left with 12 good
candidate members of Corona Australis that are H$\alpha$ emitters.  Nine of
them have spectral types earlier than M6, and are classified as very low-mass
stars. The remaining  three candidates have estimated spectral types M7-M8.5,
corresponding to brown dwarfs. This classification is summarized in
Table~\ref{tab:wfiphot}. We also include in this table the star CrA~466, which
has a mid-infrared excess but no clear H$\alpha$ emission.

	In Chamaeleon~II, ChaII~376, our only H$\alpha$ emitter, seems to be
more likely a foreground star. Still, the pair formed by ChaII~304 and
ChaII~305 could contain one or two objects in the transition from brown dwarfs
to planetary mass objects.

%########################################################################
\section{Properties of the new very low-mass members of Corona Australis} 
\label{sec:cradisc}

%______________________________________________________________
\subsection{Spatial distribution and visual binaries}\label{sec:distcra}

%_____________________________________________________________
   \begin{figure}[t]
   \centering
  \includegraphics[width=6.5cm, angle=-90, bb= 50 70 600 775]{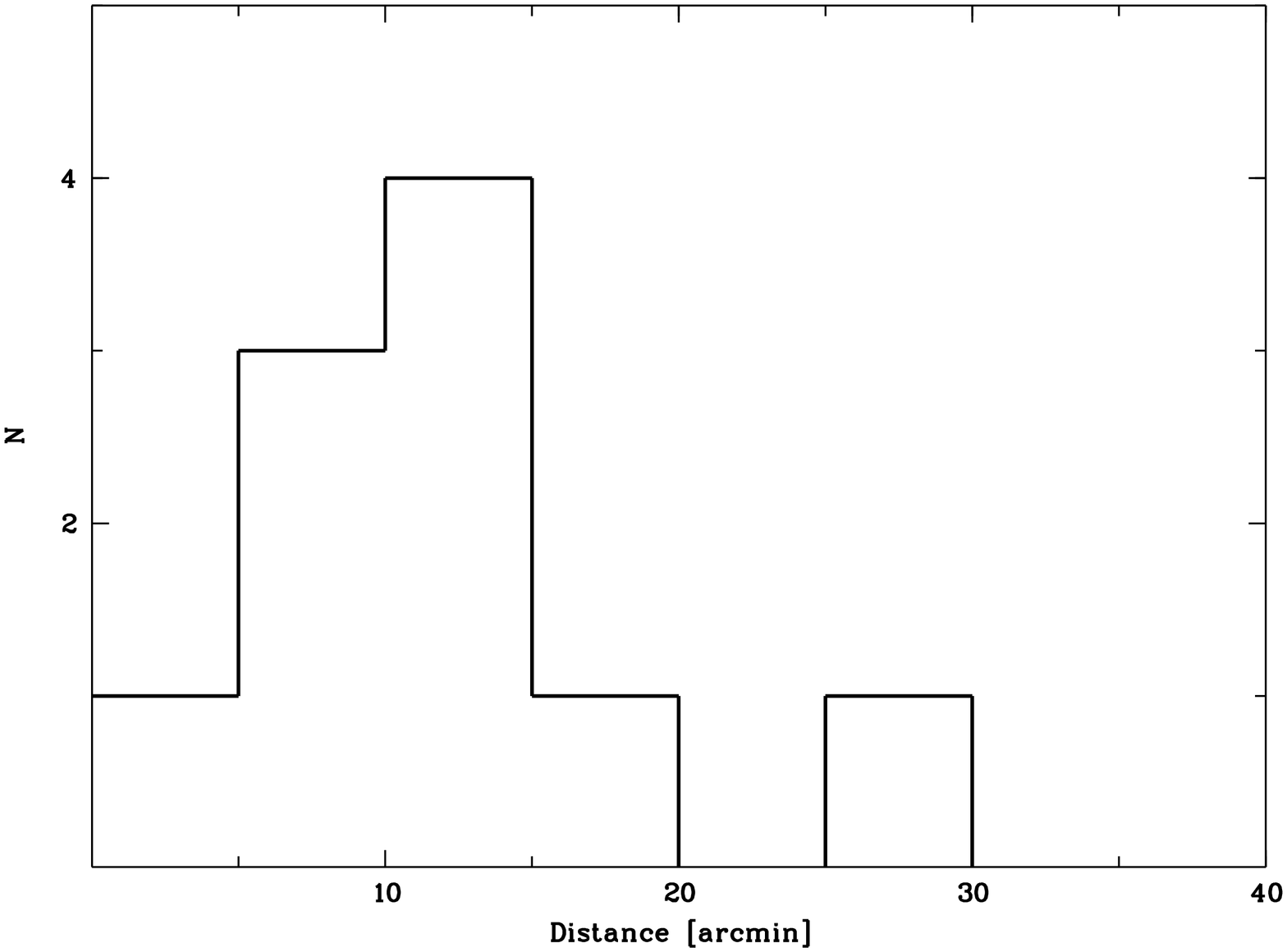}\hfill
  \includegraphics[width=6.5cm, angle=-90, bb= 50 70 600 775]{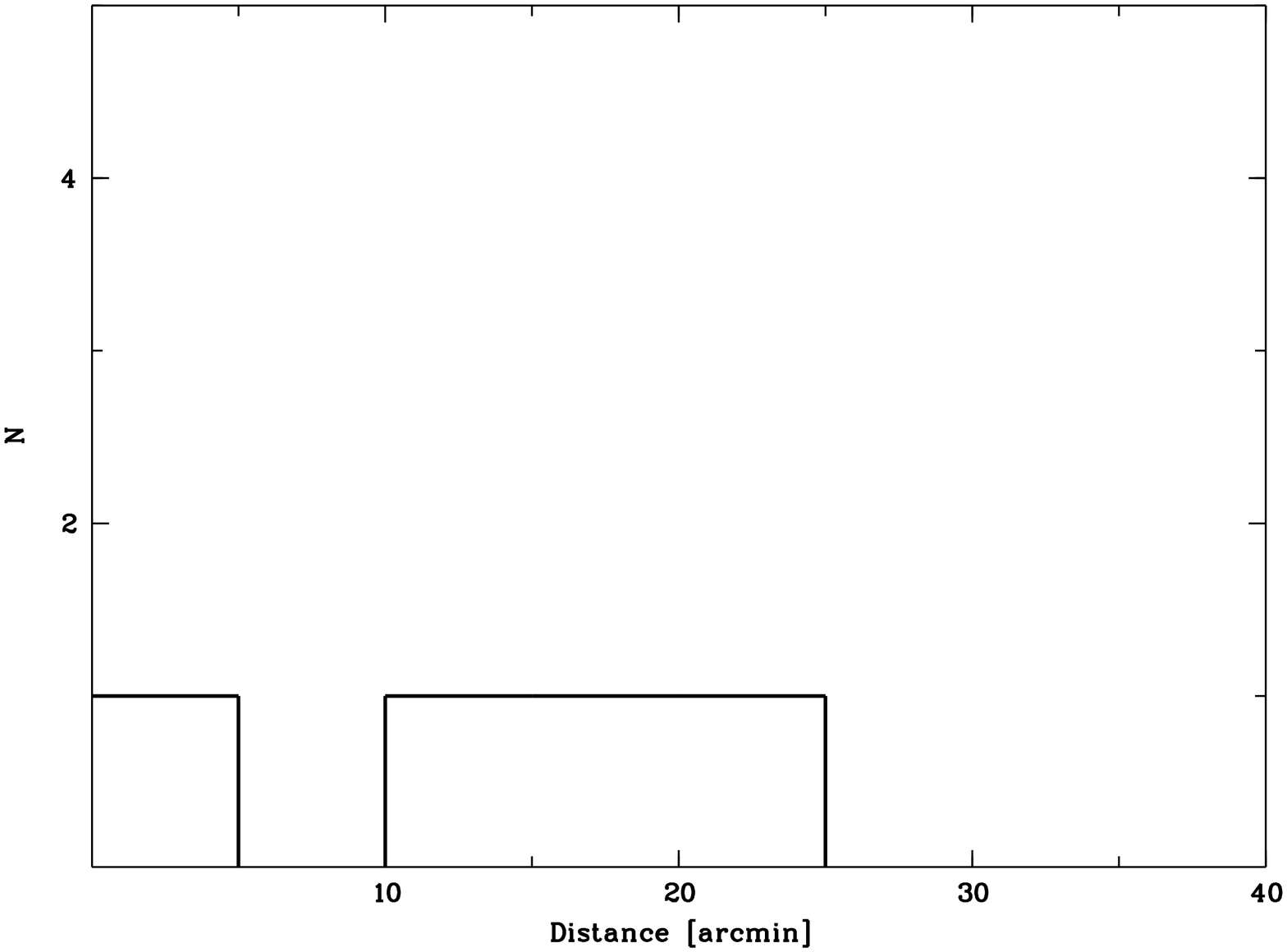}\hfill
      \caption{\footnotesize
	Distribution of the very low-mass Corona Australis stars (upper panel)
	and brown dwarf candidates (lower panel) with distance to 
	the intermediate mass star R~CrA.}
	\label{fig:distrcra}
   \end{figure}

%_____________________________________________________________
   \begin{figure}[t]
   \centering
  \includegraphics[width=8cm, angle=0]{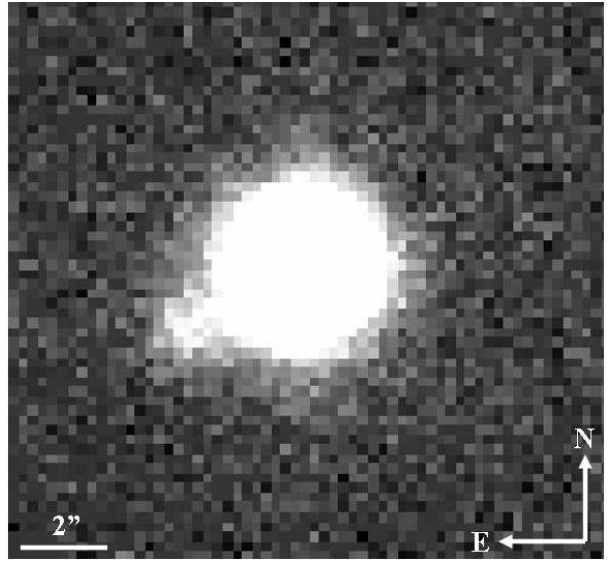}
      \caption{\footnotesize
	       The brown dwarf candidate CrA~444 (estimated spectral type 
	       M8.5) could have a very close companion. This object is seen 
	       only in our long $I$-band exposure, but was not found by the
	       automatic search.}
         \label{fig:cra444}
   \end{figure}

	As seen in Fig.~\ref{fig:crafield}, all but one of our new Corona
Australis members are located within our northern WFI field, CrA-4. This is
not surprising, given that this field contains the cloud core where most of
the previously known members of this star forming region are located (the
Coronet cluster). 

	The objects tend to be clustered around the star R~CrA. In
Fig.~\ref{fig:distrcra}, we show the number of our objects found at different
distance bins from this star. There are no obvious differences in the spatial
distribution of low-mass stars and brown dwarf candidates. Moreover, similar to
the case of Lupus~3 (\cite{lm05}), most of our new Corona Australis members are
placed at distances 5-15$\arcmin$ from the intermediate-mass star. Note that
objects very close to R~CrA might be blended by this brighter star and its
associated reflection nebula. 

	Given the low  number of objects, we cannot use the observed
distribution to test the proposed formation scenarios. We remark, however, that
this distribution suggests a connection between the very low-mass objects 
and the cloud core containing the star R~CrA, in agreement with the results in
Chamaeleon~I (cf. \cite{lm04}) and Lupus~3 (cf. \cite{lm05}).

	We also checked for possible binary systems among our new Corona
Australis members. Within a maximum estimated radius of about
9$^{\prime\prime}$  (about 1170~AU,  calculated, as in \cite{lm04} and
\cite{lm05}, by estimating the mean object surface density in our surveyed
region), we do not find any visual pair among our objects. Nonetheless, several
objects have faint neighbours that were not considered candidates according to
our selection criteria, because their colours did not match those expected for
young very low-mass objects. More interesting, the brown dwarf candidate
CrA~444 could have a faint companion at about 3$\arcsec$ (390~AU) that
was not found by the automatic search (see Fig.~\ref{fig:cra444}). This object,
\object{CrA~444b}, is only visible in our deep $I$ exposure. 

	We made an attempt to measure CrA~444b using a small aperture to remove
as much of its brighter neightbour as possible, and then performing a
correction based on other nearby objects of similar  brightness to get an
approximate instrumental $I$ magnitude. The obtained value is about 4
magnitudes fainter than the instrumental $I$ magnitude for CrA~444. In absolute
photometry, CrA~444b must have $I\sim$19~mag. Given that this object is not
seen in the $R$-band images, its $R$ magnitude must lie below our survey
completeness limit, $R\lesssim$21~mag. The optical colour of CrA~444b seems
thus to be similar to that of ChaII~304 and ChaII~305 in Chamaeleon~II. If it
belonged to the cloud, the companion to CrA~444 would have a mass below the
deuterium burning limit ($\sim 0.013\,$M$_{\odot}$).

%______________________________________________________________
\subsection{Infrared properties}\label{sec:irpropcra}

	As seen in Table~\ref{tab:irphot}, most of the ISOCAM detected objects
are very faint and visible only at 6.7 $\mu$m. Only two objects, CrA~4107 and
CrA~466, are detected in both bands. Both of them have a mid-infrared excess.

	CrA~4107 (ISO-CrA~177) is a  star (M4.5) with a very blue H$\alpha-R$
colour (--1.1), implying strong H$\alpha$ emission. These properties strongly
suggest that CrA~4107 is young and has an accretion disk. A similar correlation
between H$\alpha$ emission (detected as a blue H$\alpha-R$ colour) and a
mid-infrared excess had also been found in several Chamaeleon~I very low-mass
objects (\cite{lm04}).

	CrA~466 (ISO-CrA~127) is the only object without H$\alpha$ emission in
the sample, although this classification might be ambiguous, because its
H$\alpha-R$ colour (--0.16) would place it among the emitters if it were a
little bit brighter. Curiously, this is also the brightest mid-infrared source.
In the ($J-H$, $H-K$) diagram, CrA~466 is found in the centre of the reddening
band with an extinction value of about 8~mag. However, this is not a
particularly faint source either in the optical (I=16.01~mag) nor in the
near-infrared ($J=$\,12.834~mag, $K=$\,10.453~mag). An independent estimation
of the extinction towards this object from its $R-I$ colour and its estimated
spectral type (M4.5) yields a extinction value of only A$_V\simeq$~2.15~mag.
(We used the intrinsic colours from Comer\'on et al. \cite{comeron00} in this
computation.) Hence, it seems that the near-infrared colours of CrA~466 could
be intrinsic to the object rather than caused by high extinction, which also
hints to the presence of circumstellar material and/or accretion. Follow-up
spectroscopy of this star might confirm the presence of H$\alpha$  emission.

%______________________________________________________________
\subsection{H$\alpha$ emission}\label{sec:halcra}

%_____________________________________________________________
   \begin{figure}
   \centering
  \includegraphics[width=6.5cm, angle=-90, bb= 50 70 600 775]{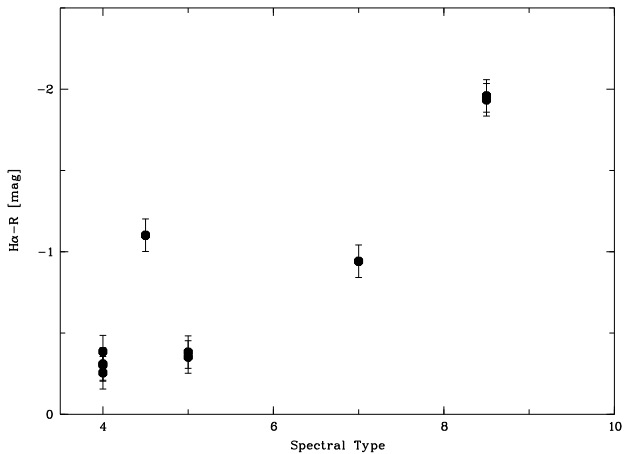}\hfill
      \caption{\footnotesize
                Measured H$\alpha-R$ colour index versus spectral type for 
      		our objects in Corona Australis. }
         \label{fig:haspt_cra}
   \end{figure}
%

%_____________________________________________________________
   \begin{figure}
   \centering
  \includegraphics[width=6.5cm, angle=-90, bb= 50 70 600 775]{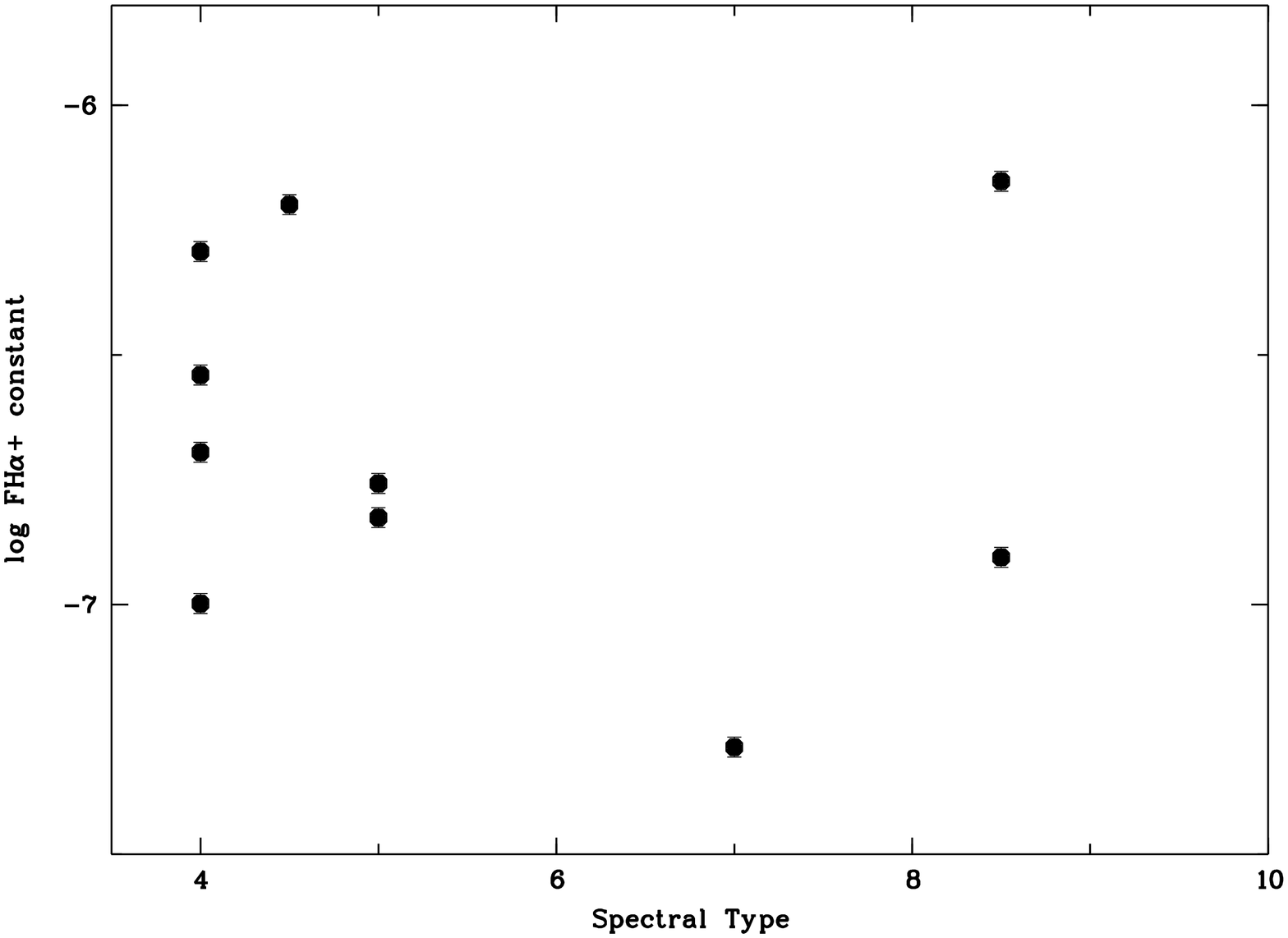}\hfill
  \includegraphics[width=6.5cm, angle=-90, bb= 50 70 600 775]{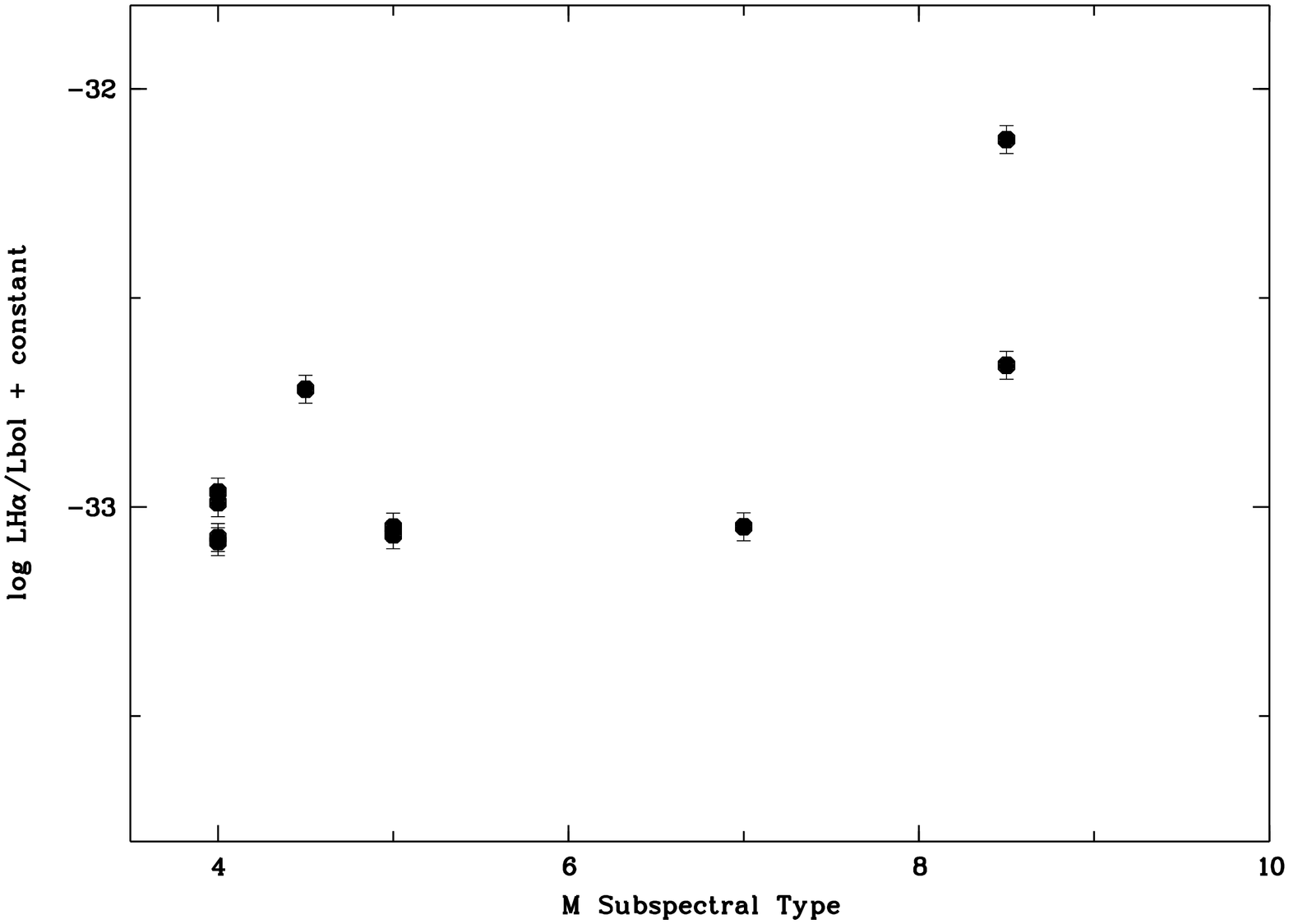}\hfill
      \caption{\footnotesize
	        \emph{Upper panel:}
      		Measured H$\alpha$ pseudoflux versus spectral type for 
      		our objects in Corona Australis.  
%		The flux clearly decreases with later spectral type.
                \emph{Lower panel:}
                Measured \~L$_{H\alpha}$/L$_{bol}$ ratio versus spectral type 
                for our objects in Corona Australis.}
         \label{fig:lhaspt_cra}
   \end{figure}

	The relation between the H$\alpha-R$ colour index and the spectral
type is shown in Fig.~\ref{fig:haspt_cra}. Like in previously studied regions 
(cf. Fig.~13 of \cite{lm04} and Fig.~7 of \cite{lm05}), an increase of emission
towards later spectral types might be present. However, given the faintness of
these later-type objects, this apparently strong emission might be just the
consequence of a lower continuum.

	To better study the emission properties of our objects, we proceeded as
in our previous works (\cite{lm04} and \cite{lm05}). First, a H$\alpha$
\emph{pseudoflux} \~F$_{\mathrm{H}\alpha}$ was computed as:

\begin{equation}\label{eq:fha}
$\~F$_{\mathrm{H}\alpha}=
\mathrm{F}_{\mathrm{H}\alpha}/\mathrm{F}_0=10^{-m_{\mathrm{H}\alpha}/2.5}
\end{equation} 

\noindent
where $m_{\mathrm{H}\alpha}$ denotes the H$\alpha$ (instrumental) magnitude,
and F$_0$ is a hypothetical H$\alpha$ absolute flux. Note that this is not the
flux in the H$\alpha$ emission line, which requires spectroscopy to be
measured. From this quantity \~F$_{\mathrm{H}\alpha}$, a H$\alpha$
\emph{pseudoluminosity} was computed using the distance estimate of 130~pc.
	
	The upper panel in Fig.~\ref{fig:lhaspt_cra} shows 
log~\~F$_{\mathrm{H}\alpha}$ versus the spectral type for our objects  in
Corona Australis. Although the values of the H$\alpha$ pseudoflux are
consistent with those measured in other star forming regions, in Corona
Australis we do not see a decrease of log~\~F$_{\mathrm{H}\alpha}$ with later
spectral type (compare the upper panel of Fig.~\ref{fig:lhaspt_cra} with the
upper panels of Fig.~14 in \cite{lm04} and Fig.~8 in \cite{lm05}). On the other
hand, no trend with the spectral type is found either in the ratio of H$\alpha$
pseudoluminosity to bolometric luminosity,
\~L$_{\mathrm{H}\alpha}$/L$_{\mathrm{bol}}$.\footnote{\footnotesize The
bolometric luminosity was computed from the I-magnitude and estimated spectral
type for each object as in Comer\'on et al. (\cite{comeron00}).} 
This is in agreement with our results in  other regions (cf. \cite{lm04} and
\cite{lm05}).

%
%########################################################################
\section{No brown dwarfs in Chamaeleon~II?}\label{sec:chaiidisc}

	Our survey in Chamaeleon~II has provided no good candidate members in
this star forming region. Our only H$\alpha$ emitter has been almost completely
discarded as a young very low-mass object, because its inferred mass and
spectral type from optical and near-infrared photometry are not consistent with
each other. It seems also unlikely that the objects without H$\alpha$ emission
detected by 2MASS are members of Chamaeleon~II, since they do not exhibit
appreciable near-infrared excesses. 

	Although the surveyed area is relatively large, these results do not
automatically imply a lack of brown dwarfs in Chamaeleon~II. Not all young
objects show H$\alpha$ emission or near-infrared excess; thus, it cannot be
excluded that some of the objects in our candidate list might eventually turn
out to be young members of this dark cloud. Furthermore, it must be taken into
account that Chamaeleon~II is a very large region: It has an extension of about
1.5 deg$^2$ and contains eleven cores according to C$^{18}$O measurements
(Mizuno et al. \cite{mizuno99}). Our field ChaII-3 covers only part of one of
these cores. Moreover, given the large and variable extinction in this region,
objects deeply embedded in the cloud might not be detectable at optical
passbands. Hence, the apparent lack of brown dwarfs inferred from our survey
might be due to selection effects.

	Another thing to keep in mind is that very few T~Tauri stars (the
higher mass counterparts to the young brown dwarfs) are known in Chamaeleon~II.
The brown dwarf population identified so far in the neighbouring Chamaeleon~I
cloud is about 10\% of its stellar population (e.g. Comer\'on et al.
\cite{comeron00}; \cite{lm04}; Luhman \cite{luhman04}). If the same ratio held
also for Chamaeleon~II, we would expect to find only two brown dwarfs in this
dark cloud.

%########################################################################
\section{Conclusions} \label{sec:concl}

	We have presented the results of a deep multi-band survey in the Corona
Australis and Chamaeleon~II star forming regions. Low-mass member candidates
were selected from $RI$ and H$\alpha$ photometry, and spectral types were
assigned by means of their $M855$--$M915$ colours. 

	In Corona Australis, we identified  ten stars and  three brown dwarf
candidates in an area of about 0.6~deg$^2$, mostly located in the area of the
Coronet cluster. All but one of these objects exhibit H$\alpha$ emission
according to their H$\alpha-R$ colours. The only non-emitter included in our
final sample, CrA~466, is the probable counterpart of an ISOCAM source with
mid-infrared excess. Its H$\alpha$ emission is not clear in our data due to the
photometric errors. Apart from this low-mass star, other four objects from our
candidate list had been detected by ISOCAM, but only one, CrA~4107, shows a
mid-infrared excess. This M4.5 object is also a strong H$\alpha$ emitter.
Hence, it is most probably surrounded by an accretion disk.

	There is no obvious difference in the spatial distribution of stars and
brown dwarf candidates: They all tend to be clustered around the intermediate
mass star R~CrA. However, the number of objects is too low to drive any
significant conclusions about their formation process from this distribution.

	The visual binary frequency of our Corona Australis objects is very
low: None of our candidate members has a close neighbour with colours
suggesting membership of the star forming region. The only exception is the
brown dwarf candidate CrA~444, which has a faint close companion in our deep
$I$ exposure. More observations are needed to certify whether this is a real
double system. 

	In Chamaeleon~II, an area of about 0.6~deg$^2$ was observed to the
North-West of the cloud. We do not find any good candidate members in this
region. Our only H$\alpha$ emitter, ChaII~376, seems to be a foreground star. A
pair of very faint objects, ChaII~304 and ChaII~305, are identified with a
near-infrared source from the 2MASS survey. The colours of these objects are
consistent with two low-mass brown dwarfs or planetary mass objects close to 
the deuterium burning limit. However, they need spectral confirmation of their
true nature. The rest of objects could be extincted young Chamaeleon~II members
or older field stars. Since none of them shows H$\alpha$ emission or an
infrared excess, nothing can be stated until spectroscopy is available for
them.

	Given these results and the uncertainties in our analysis, we are led
to the conservative conclusion that it is unlikely that the identified
candidates are substellar members of Chamaeleon~II. Hence, our data seem to
indicate a lack of brown dwarfs in this dark cloud. We cannot exclude, however,
that this result is a consequence of having surveyed only a limited region. The
Chamaeleon~II dark cloud has an extension of about 1.5 deg$^2$ and contains
eleven cores according to C$^{18}$O measurements (Mizuno et al.
\cite{mizuno99}). Moreover, deeply embedded brown dwarfs might not be detected
in optical passbands. Larger and deeper surveys are needed to state if the
deficit of brown dwarfs observed here is real.

%########################################################################
\begin{acknowledgements}
We acknowledge much support from I. Baraffe and F. Allard, providing us with
synthetic photometry from their brown dwarf models. We are also very grateful
to C.~Bailer-Jones for his help with the field selection, and to A.~Scholz for
useful discussions. J.~E. kindly thanks E.~Pompei and the 2p2-team for their
support during the observations.

This publication makes use of data products from the Two Micron All Sky Survey
(2MASS), a joint product of the University of  Massachussets and the Infrared
Processing and Analysis centre/California  Institute of Technology, funded by
the US National Aeronautics and Space  Administration and the US National
Science Foundation. We used the SIMBAD database and the Vizier catalogue
service, both operated at the \emph{Centre de Donn\'ees astronomiques de
Strasbourg (CDS)} in Strasbourg (France), as well as the NASA/IPAC Infrared
Science Archive, operated by the Jet Propulsion Laboratory, California
Institute of Technology, under contract with the US National Aeronautics and
Space Administration.

This work was supported by the German 
\emph{Deut\-sche Forschungs\-ge\-mein\-schaft (DFG)}, projects EI\,409/7-1 and
EI\,409/7-2. B.\,L.\,M. acknowledges financial support from the Spanish
Ministerio de Educaci\'on y Ciencia through a \emph{Juan de la Cierva} 
fellowship.

\end{acknowledgements}

%########################################################################


\begin{thebibliography}{}

\bibitem[2002]{alcala02} 
Alcal\'a, J.~M., Radovich, M., Silvotti, R. et al. 2002,  \procspie, 4836, 406 
	
\bibitem[2002]{apai02}
Apai, D., Pascucci, I., Henning, Th. et al. 2002, \apj \ 573, L115

\bibitem[1998]{baraffe98} 
Baraffe, I., Chabrier, G., Allard, F., \& Hauschildt, P.~H.\ 1998, 
\aap \ 337, 403 

\bibitem[2001]{barrado01}
Barrado y Navascu\'es, D., Stauffer, J., Brice\~no, C. et al. 2001, 
\apjs \ 134, 103

\bibitem[2004a]{barrado04a}
Barrado y Navascu\'es, D. Mohanty, S. \& Jayawardhana, R. 2004a, 
\apj \ 604, 248


\bibitem[2004b]{barrado04b}
Barrado y Navascu\'es, D. \& Jayawardhana, R. 2004b, \apj \ 615, 840

\bibitem[2003]{bate03} 
Bate, M.~R., Bonnell, I.~A. \& Bromm, V. 2003, \mnras \ 339, 557

\bibitem[1999]{bejar99} 
B{\' e}jar, V.~J.~S., Zapatero Osorio, M.~R. \& Rebolo, R.\ 1999, 
\apj \ 521, 671 

\bibitem[1996]{bertin96}
Bertin, E. \& Arnouts, S. 1996, \aaps \ 117, 393

\bibitem[1988]{bessell88}
Bessell, M.~S. \& Brett, J.~M. 1988, \pasp \ 100, 1134

\bibitem[2000]{chabrier00} 
Chabrier, G., Baraffe, I., Allard, F., \& Hauschildt, P.\ 2000, \apj \ 542, 464

\bibitem[2000]{comeron00}
Comer\'on, F., Neuh\"auser, R. \& Kaas, A.~A. 2000, \aap \ 359, 269

\bibitem[2003]{comeron03} 
Comer{\' o}n, F., Fern\'andez, M., Baraffe, I. et al. \ 2003, \aap \ 406, 1001

\bibitem[2004]{comeron04} 
Comer{\' o}n, F., Reipurth, B., Henry, A. \& Fern\'andez, M. \ 2004, 
\aap \ 417, 583
	
\bibitem[2003]{delgado03} 
Delgado Donate, E., Clarke, C.~J. \& Bate, M.~R. 2003, \mnras \ 342, 926

\bibitem[2001]{fernandez01} 
Fern\'andez, M. \& Comer\'on, F. 2001, \aap \ 380, 264

\bibitem[1992]{gauvin92} 
Gauvin, L.~S.~\& Strom, K.~M.\ 1992, \apj \ 385, 217 
	
\bibitem[1991]{graham91} 
Graham, J.~A. 1991, in: 
Reipurth, B. (Ed.), ESO Scientific Report No.11,  
\emph{Low Mass Star Formation in Southern Molecular Clouds}, 185
	
\bibitem[1991]{hughes91} 
Hughes, J.~D., Hartigan, P., Graham, J.~A. et al. 1991, \apj \ 101, 1013
	
\bibitem[1992]{hughes92}
Hughes, J.~D. \& Hartigan, P. 1992, \aj \ 104, 680

\bibitem[2002]{jay02}
Jayawardhana, R., Mohanty, S. \& Basri, G. 2002, \apj \ 578, L141

\bibitem[2003]{jay03}
Jayawardhana, R., Mohanty, S. \& Basri, G. 2003, \apj \ 592, 282

\bibitem[1991]{kirkpatrick91}
Kirkpatrick, J.~D., Henry, T.~J. \& McCarthy, D.~W. 1991, \apjs \ 77, 417 

\bibitem[2003]{kroupa03}
Kroupa, P. \& Bouvier, J. 2003, \mnras \ 346, 369

\bibitem[2005]{lamm05} 
Lamm, M.~H., Mundt, R., Bailer-Jones, C.~A.~L. \& Herbst, W. 2005, 
\aap \ 430, 1005

\bibitem[1992]{landolt92} 
Landolt, A.~U.\ 1992, \aj \ 104, 340 

\bibitem[Paper~1]{lm04} 
L\'opez Mart\'{\i}, B., Eisl\"offel, J., Scholz, A. \& Mundt, R. \ 2004,
\aap \ 416, 555 (Paper~1)

\bibitem[Paper~2]{lm05} 
L\'opez Mart\'{\i}, B., Eisl\"offel, J., \& Mundt, R. \ 2005, \aap , in press 
(Paper~2)
	
\bibitem[1979]{loren79} 
Loren, R.~B. 1979, \apj \ 227, 832

\bibitem[1999]{luhman99} 
Luhman, K.~L. 1999, \apj \ 525, 466 

\bibitem[2004a]{luhman04} 
Luhman, K.~L. 2004, \apj \ 602, 816

\bibitem[1981]{marraco81} 
Marraco, H.~G. \& Rydgren, A.~E. 1981, \aj \ 86, 62

\bibitem[2001]{martin01} 
Mart{\' i}n, E.~L., Dougados, C., Magnier, E., et al. 2001, \apjl \ 561, L195 

\bibitem[1999]{mizuno99} 
Mizuno, A., Hayakawa, T., Tachihara, K.,~et al.\ 1999, \pasj \ 51, 859 

\bibitem[2004]{mohanty04}
Mohanty, S., Jayawardhana, R., Natta, A. et al. 2004, \apj \ 609, L33

\bibitem[2005]{mohanty05}
Mohanty, S., Jayawardhana, R. \& Basri, G. 2005, \apj \ 626, 498

\bibitem[2001]{natta01} 
Natta, A.~\& Testi, L.\ 2001, \aap \ 376, L22 

\bibitem[2000]{neuhauser00}
Neuh\"auser, R., Walter, F.~M., Covino, E. et al. 2000, A\&ASS \ 146, 323

\bibitem[1999]{olofsson99} 
Olofsson, G., Huldtgren, M., Kaas, A.~A. et al.\ 1999, \aap, \ 350, 883 
	
\bibitem[2002]{padoan02} 
Padoan, P. \& Nordlund, \AA \ 2002, \apj \ 576, 870

\bibitem[2003]{pascucci03}
Pascucci, I., Apai, D., Henning, Th. et al. 2003, \apj 590, L111
	
\bibitem[1998]{patten98} 
Patten, B. 1998, in:
Donahue, R.~A. \& Bookbinder, J.~A. (Eds.):
\emph{The Tenth Cambridge Workshop on Cool Stars}, ASP Conf. Ser. 125, 34

\bibitem[2003]{persi03} 
Persi, P., Marenzi, A.~R., G{\' o}mez, M., \& Olofsson, G.\ 2003, 
\aap, 399, 995 
 
\bibitem[2001]{reipurth01} 
Reipurth, B.~\& Clarke, C.\ 2001, \aj \ 122, 432 

\bibitem[1985]{rieke85}
Rieke, G.~K. \& Lebofsky, M.~J. 1985, \apj \ 288, 618

\bibitem[2004]{scholz04} 
Scholz, A. \& Eisl\"offel, J. 2004, \aap \ 429, 1007

\bibitem[1977]{schwartz77} 
Schwartz, R.~D. 1977, \apjs \ 35, 161
	
\bibitem[1991]{schwartz91} 
Schwartz, R.~D. 1991, in 
ESO Scientific Report No.11,  
\emph{Low Mass Star Formation in Southern Molecular Clouds}, 
Ed. Reipurth, B., 93

\bibitem[2004]{sterzik04} 
Sterzik, M.~F., Pascucci, I., Apai, D., et al. 2004, \aap \ 427, 245 
	
\bibitem[1987]{stetson87}
Stetson, P.~B. 1987, \pasp \ 99, 191
	
\bibitem[1984]{taylor84}
Taylor, K.~N.~R. \& Storey, J.~W.~V. 1984, \mnras \ 209, 5P

\bibitem[2002]{testi02}
Testi, L., Natta, A., Comer\'on, F. et al. 2002, \aap \ 393, 597

\bibitem[2001]{vuong01} 
Vuong, M.H., Cambr\'esy, L. \& Epchtein, N., 2001, \aap \ 379, 208
	
\bibitem[1997]{walter97} 
Walter, F.~M., Vrba, F.~J., Wolk, S.~J. et al. 1997, \aj \ 114, 1544 
	
\bibitem[2004]{wang04} 
Wang, H., Mundt, R., Henning, Th. \& Apai, D. 2004, \apj \ 617, 1191
	
\bibitem[1997]{whittet97} 
Whittet, D.~C.~B., Prusti, T., Franco, G.~A. et al.\ 1997, \aap \ 327, 1194 

\bibitem[2004]{whitworth04} 
Whitworth, A.~P. \& Zinnecker, H. 2004, astro-ph/0408522 

\bibitem[1992]{wilking92} 
Wilking, B.~A., Greene, T.~P., Lada, C.~J. et al 1992, \apj \ 397, 520
	
\bibitem[1997]{wilking97} 
Wilking, B.~A., Mc Caughrean, M.~J., Burton, M.~G. et al 1997, \aj \ 114, 2029
	
\bibitem[1999]{wilking99} 
Wilking, B.~A., Greene, T.~P. \& Meyer, M.~L. 1999, \aj \ 117, 469

\bibitem[2005]{young05}
Young, K.~E., Harvey, P.~M., Brooke, T.~Y. et al. 2005, astro-ph/0503670

\end{thebibliography}
\end{document}